\documentclass[preprintnumbers,prd,showpacs,floatfix,superscriptaddress,nofootinbib,twocolumn, letterpaper]{revtex4-1}
\pdfoutput=1

\usepackage{longtable}
\usepackage{bm}
\usepackage{relsize}
\usepackage{amsfonts}
\usepackage{amsmath}
\usepackage{amssymb,epsf}
\usepackage{latexsym}
\usepackage{graphicx,epsfig}
\usepackage{amssymb}
\usepackage{float}
\usepackage{subfigure}
\usepackage{epstopdf}
\usepackage[colorlinks=true,citecolor=blue,linkcolor=blue,urlcolor=black]{hyperref}
\usepackage{dcolumn}
\usepackage{psfrag}
\usepackage{wrapfig}
\usepackage{makeidx}
\usepackage{epsf}
\usepackage{color}
\usepackage{multirow}
\usepackage{mathtools}

\begin{document}

\title{Non-exotic traversable wormhole solutions in linear $f\left(R,T\right)$ gravity}

\author{Jo\~{a}o Lu\'{i}s Rosa}
\email{joaoluis92@gmail.com}
\affiliation{Institute of Physics, University of Tartu, W. Ostwaldi 1, 50411 Tartu, Estonia}

\author{Paul Martin Kull}
\email{paulmartin831@hotmail.com}
\affiliation{Institute of Physics, University of Tartu, W. Ostwaldi 1, 50411 Tartu, Estonia}

\date{\today}

\begin{abstract} 
In this work we analyze traversable wormhole solutions in the linear form of $f\left(R,T\right)=R+\lambda T$ gravity satisfying the Null, Weak, Strong, and Dominant Energy Conditions (NEC, WEC, SEC, and DEC respectively) for the entire spacetime. These solutions are obtained via a fully analytical parameter space analysis of the free parameters of the wormhole model, namely the exponents controlling the degree of the redshift and shape functions, the radius of the wormhole throat $r_0$, the value of the redshift function at the throat $\zeta_0$, and the coupling parameter $\lambda$. Bounds on these free parameters for which the energy conditions are satisfied for the entire spacetime are deduced and two explicit solutions are provided. Even if some of these bounds are violated, leading to the violation of the NEC at some critical radius $r_c>r_0$, it is still possible to find physically relevant wormhole solutions via a matching with an exterior vacuum spacetime in the region where the energy conditions are still satisfied. For this purpose, we deduce the set of junction conditions for the form of $f\left(R,T\right)$ considered and provide an explicit example. These results seem to indicate that a wide variety of non-exotic wormhole solutions are attainable in the $f\left(R,T\right)$ theory without the requirement of fine-tuning.
\end{abstract}


\maketitle

\section{Introduction}\label{sec:intro}

A wormhole is a topological object connecting two spacetime manifolds. In the theory of General Relativity (GR), several wormhole solutions connecting asymptotically flat \cite{morris1,visser1,visser2,visser3} and asymptotically (anti-)de-Sitter \cite{lemos1} spacetimes have been obtained. However, these solutions feature a major drawback: in GR, the geometrical condition necessary for a wormhole spacetime to be traversable, known as the \textit{flaring-out} condition, is incompatible with the Null Energy Condition (NEC), a condition that states that any null observer should measure a non-negative average energy density in the spacetime. When the matter components violate the NEC, the matter is denoted as \textit{exotic}. In the pursuit of physically relevant wormhole solutions, one must thus recur to modified theories of gravity.

The literature concerning wormhole solutions in modified theories of gravity is quite extensive \cite{agnese1,nandi1,bronnikov1,camera1,camera2,lobo1,garattini1,lobo2,garattini2,lobo3,garattini3,myrzakulov1} (we refer the reader to Ref.\cite{lobo4} for a review). In this context, the higher-order curvature terms are the ones responsible for maintaining the geometry of the wormhole throat, while the matter components are kept non-exotic. This result can be achieved in multiple frameworks, for example $f\left(R\right)$ gravity \cite{lobo5}, non-minimal couplings \cite{garcia1,garcia2}, additional fundamental fields \cite{harko1}, Einstein-Gauss-Bonnet gravity \cite{bhawal1,dotti1,mehdizadeh1}, Brans-Dicke gravity \cite{anchordoqui1}, braneworld configurations \cite{lobo6}, and hybrid metric-Palatini gravity \cite{capozziello1,rosa1,rosa2,rosalol}.

In this work, we are particularly interested in an extension of $f\left(R\right)$ gravity known as $f\left(R,T\right)$ gravity, where $R$ is the Ricci scalar and $T$ is the trace of the stress-energy tensor \cite{harko2}. This theory has been explored in a wide variety of topics including dark matter models \cite{zaregonbadi1}, compact objects including white dwarfs and exotic solutions \cite{dey1,carvalho1,deb1,maurya1,bhatti1}, cosmological solutions including reconstruction methods \cite{velten1,mirza1,houndjo1, houndjo2,jamil1}, stability analyses \cite{alvarenga1}, the Palatini formulation \cite{wu1} and junction conditions \cite{rosa3, rosa4}. Recently, this theory was also shown to provide relevant solutions for wormhole spacetimes \cite{dixit1,banerjee1,mishra1,sahoo1,moraes1}. The $f\left(R,T\right)$ is currently one of the most actively studied modified theories of gravity. Indeed, the recent derivation of an alternative scalar-tensor representation of the theory \cite{rosa3} has opened a new research branch with particular emphasis on cosmology \cite{goncalves1,goncalves2,pinto1} and braneworld scenarios \cite{bazeia1,rosa5,rosa6,rosa7}.

In the majority of the wormhole works mentioned before, even though the higher-order curvature terms can provide solutions satisfying the NEC at the throat, frequently this condition is violated elsewhere, thus compromising the physical relevance of the solution. A possible resolution for this problem is to perform a matching with an exterior vacuum spacetime in the region where the NEC is satisfied, effectively replacing the problematic domain of the solution with an infinitely thin shell, a usually tedious method that frequently involves fine-tuning. Furthermore, even in the few papers where this procedure is not necessary, the wormhole solutions satisfying the NEC for the whole spacetime are frequently obtained through a trial-and-error method, and no clear analytical study of the parameter space is provided. The first goal of this work is to fulfil this gap by performing a fully analytical study of the parameter space of wormhole spacetimes in a linear model of $f\left(R,T\right)$ gravity. Furthermore, we also aim to extend the physical relevance of the solutions obtained by forcing not only the NEC to be satisfied, but also the Weak, Strong, and Dominant Energy Conditions (WEC, SEC, and DEC, respectively).

This paper is organized as follows. In Sec.\ref{sec:theory} we introduce the $f\left(R,T\right)$ gravity in its usual geometrical representation, and also outline general considerations about wormhole spacetimes; in Sec.\ref{sec:solutions} we perform an analytical study of the parameter space of the model considered in order to obtain the necessary parameter bounds for the energy conditions to be satisfied, and provide a couple of examples of solutions; in Sec.\ref{sec:matching} we derive the junction conditions of the model considered and outline how to obtain a physically relevant wormhole solution when one of the parameter bounds previously obtained is violated; and in Sec.\ref{sec:concl} we trace our conclusions. Part of this work, namely Sec. \ref{sec:matching}, is included in P.M.K. BSc thesis \cite{kullthesis}.

\section{Theory and framework}\label{sec:theory}

\subsection{Action and equations of the $f\left(R,T\right)$ gravity}\label{sec:geom}

The $f\left(R,T\right)$ theory of gravity is described by an action function $S$ of the form
\begin{equation}\label{eq:geo_action}
    S = \frac{1}{2\kappa^2} \int_{\Omega}\sqrt{-g} f\left(R,T\right) d^4 x+ \int_{\Omega} \sqrt{-g} \mathcal{L}_m d^4 x,
\end{equation}
where $\kappa^2\equiv8\pi G/c^4$, where $G$ is the gravitational constant and $c$ is the speed of light, $\Omega$ is a 4-dimensional spacetime manifold on which one defines a set of coordinates $x^\mu$, $g_{\mu\nu}$ is the metric tensor with positive signature written in terms of the coordinates $x^\mu$ and with a determinant $g$, $f\left(R,T\right)$ is an arbitrary well-behaved function of the Ricci scalar $R=g^{\mu\nu}R_{\mu\nu}$, where $R_{\mu\nu}$ is the Ricci tensor, and the trace of the stress-energy tensor $T=g^{\mu\nu}T_{\mu\nu}$. The stress-energy tensor $T_{\mu\nu}$ is defined in terms of the variation of the matter Lagrangian $\mathcal L_m$ with respect to the metric as
\begin{equation}\label{eq:geo_def_tab}
T_{\mu\nu}=-\frac{2}{\sqrt{-g}}\frac{\delta\left(\sqrt{-g}\mathcal L_m\right)}{\delta g^{\mu\nu}}.
\end{equation}
To ease the notation, we shall adopt a system of geometrized units for which $G=c=1$, and hence $\kappa^2=8\pi$. 

A variation of Eq.\eqref{eq:geo_action} with respect to the metric $g_{\mu\nu}$ leads to the modified field equations of the $f\left(R,T\right)$ gravity in the form
\begin{equation}\label{eq:geo_fieldeq}
\begin{multlined}
    f_R R_{\mu\nu}-\frac{1}{2}g_{\mu\nu}f\left(R,T\right) + \left(g_{\mu\nu}\square-\nabla_\mu\nabla_\nu\right)f_R \\ 
    = 8\pi T_{\mu\nu}-f_T (T_{\mu\nu}+\Theta_{\mu\nu}),
\end{multlined}
\end{equation}
where we have defined the partial derivatives of $f$ as $f_R\equiv\partial f/\partial R$ and $f_T\equiv\partial f/\partial T$, $\nabla_\mu$ and $\square\equiv\nabla^\sigma\nabla_\sigma$ are the covariant derivative and the D'Alembert operator defined in terms of the metric $g_{\mu\nu}$, and $\Theta_{\mu\nu}$ is an auxiliary tensor defined in terms of the variation of $T_{\mu\nu}$ as
\begin{equation}\label{eq:geo_def_theta}
    \Theta_{\mu\nu}\equiv g^{\rho\sigma}\frac{\delta T_{\rho\sigma}}{\delta g^{\mu\nu}}.
\end{equation}
The explicit form of the tensor $\Theta_{\mu\nu}$ will be defined upon the specification of a matter Lagrangian $\mathcal L_m$ or, equivalently, a stress-energy tensor $T_{\mu\nu}$.

In this work, we work with a function $f\left(R,T\right)$ of the form $f\left(R,T\right)=R+\lambda T$ . This is one of the most commonly used forms of the function $f\left(R,T\right)$ in the literature as it provides a simple extension of GR where the action $S$ depends linearly on $T$. The equations of motion in Eq.\eqref{eq:geo_fieldeq} in this particular case simplify to
\begin{equation}\label{eq:fieldeq}
R_{\mu\nu}-\frac{1}{2}g_{\mu\nu}\left(R+\lambda T\right)=8\pi T_{\mu\nu}-\lambda\left(T_{\mu\nu}+\Theta_{\mu\nu}\right).
\end{equation}
In the following section, we will derive traversable wormhole solutions of the field equations above.

\subsection{Traversable wormhole spacetimes}\label{sec:ansatz}

In this work we focus on static and spherically symmetric wormhole solutions. The metric that describes a general static and spherically symmetric spacetime can be written in the usual set of spherical coordinates $\left(t,r,\theta,\phi\right)$ as
\begin{equation}\label{eq:metric}
ds^2=-e^{\zeta\left(r\right)}dt^2+\left[1-\frac{b\left(r\right)}{r}\right]^{-1}dr^2+r^2d\Omega^2,
\end{equation}
where $\zeta\left(r\right)$ is the redshift function, $b\left(r\right)$ is the shape function, and $d\Omega^2=d\theta^2+\sin^2\theta d\phi^2$ is the surface-element on the two-sphere. For the wormhole to be traversable, the functions $\zeta\left(r\right)$ and $b\left(r\right)$ must satisfy a few conditions. First, the spacetime must not feature any event horizons, as to allow a traveller to cross the throat of the wormhole at $r=r_0$ and still be able to escape the interior region. For this requirement to be fulfilled, the redshift function must remain finite throughout the entire spacetime, i.e., $|\zeta\left(r\right)|<\infty$. The second condition, also known as the flaring-out condition, is a geometrical condition at the wormhole throat that can be expressed in the form of two boundary conditions on the shape function as
\begin{equation}\label{eq:flaring_out}
b\left(r_0\right)=r_0,\qquad b'\left(r_0\right)<1.
\end{equation}
In the literature, a wide variety of forms for both the redshift and the shape functions have been explored. In this work, we focus our analysis on the following two families of functions that satisfy the requirements described above:
\begin{equation}\label{eq:def_zeta}
\zeta\left(r\right)=\zeta_0\left(\frac{r_0}{r}\right)^\alpha,
\end{equation}
\begin{equation}\label{eq:def_shape}
b\left(r\right)=r_0\left(\frac{r_0}{r}\right)^\beta,
\end{equation}
where $\zeta_0$ is an arbitrary constant and the parameters $\alpha$ and $\beta$ are constant exponents. To guarantee the asymptotic flatness of the solution, it is necessary that $\alpha$ and $\beta$ are strictly positive. 

In what concerns the matter sector, we shall assume that matter is well described by an anisotropic perfect fluid, i.e., the stress-energy tensor can be written in the form
\begin{equation}\label{eq:matter}
T_a^b=\text{diag}\left(-\rho,p_r,p_t,p_t\right),
\end{equation}
where $\rho$ is the energy density, $p_r$ is the radial pressure, and $p_t$ is the tangential pressure. To preserve the spherical symmetry and time-independence of the solution, all matter quantities are assumed to depend solely in the radial coordinate, i.e., $\rho\equiv\rho\left(r\right)$, $p_r\equiv p_r\left(r\right)$, and $p_t\equiv p_t\left(r\right)$. The wormhole solutions will be considered of physical relevance if the matter quantities satisfy the  Null Energy Condition (NEC), i.e., $T_{ab}k^a k^b\geq 0$, for some null vector $k^a$. For a stress-energy tensor $T_a^b$ of the form given in Eq.\eqref{eq:matter}, the NEC can be translated into the following two constraints on the matter variables:
\begin{equation}\label{eq:NEC}
\rho+p_r>0,\qquad \rho+p_t>0.
\end{equation}

Within the context of general relativity, the flaring-out condition (Eq.\eqref{eq:flaring_out}) and the NEC (Eq.\eqref{eq:NEC}) are incompatible. For the flaring-out condition to be satisfied, one must have $G_{\mu\nu}k^\mu k^\nu<0$ where $G_{\mu\nu}$ is Einstein's tensor and $k^\mu$ is an arbitrary null vector. This implies, from the Einstein's field equations $G_{\mu\nu}\propto T_{\mu\nu}$, that $T_{\mu\nu}k^\mu k^\nu<0$, which upon choosing a stress-energy tensor $T_{\mu\nu}$ of the form given in Eq.\eqref{eq:matter} yields precisely the violation of Eq.\eqref{eq:NEC}. However, the scenario changes in modified theories of gravity, in which Einstein's tensor $G_{\mu\nu}$ becomes proportional to an effective stress-energy tensor $T_{\mu\nu}^{\text{eff}}$ which includes not only contributions from the matter sector but also contributions from the extra gravitational degrees of freedom. Thus, even though the effective stress-energy tensor must satisfy $T_{\mu\nu}^{\text{eff}}k^\mu k^\nu<0$ to fulfil the flaring-out condition, it is still possible for the matter stress-energy tensor to satisfy the NEC, i.e., $T_{\mu\nu}k^\mu k^\nu>0$, provided that the extra gravitational contributions compensate the positive matter contributions.

In the following sections, we will obtain wormhole solutions that not only satisfy the NEC but also the Weak, Strong, and Dominant energy conditions (WEC, SEC, and DEC, respectively). These three energy conditions are extensions of the NEC which not only require Eq. \eqref{eq:NEC} to hold, but also some extra conditions: the WEC requires a positivity of the energy density, i.e., $\rho>0$, the SEC requires that $\rho+p_r+2p_t>0$, and the DEC requires a dominance of the energy density over the pressures, i.e., $\rho>|p_r|$ and $\rho>|p_t|$. One can thus verify that both the WEC and the SEC imply the NEC, but they are independent of each other, and that the DEC implies the WEC and, consequently, the NEC.

\section{Smooth wormhole solutions}\label{sec:solutions}

Let us now obtain explicit wormhole solutions from the field equations. Taking a metric of the form given in Eq.\eqref{eq:metric} and a stress-energy tensor of the form of Eq.\eqref{eq:matter} into the modified field equations in Eq.\eqref{eq:fieldeq}, one obtains three independent field equations that can be solved with respect to the matter variables $\rho$, $p_r$ and $p_t$, which take the forms:

\begin{equation}\label{eq:fieldmatter1}
8\pi\rho=\frac{\lambda}{6}\left(p_r+2p_t-9\rho\right)+\frac{b'}{r^2},
\end{equation}
\begin{equation}\label{eq:fieldmatter2}
8\pi p_r=\frac{\lambda}{6}\left(3\rho-7p_r-2p_t\right)-\frac{b}{r^3}+\frac{\zeta'}{r}\left(1-\frac{b}{r}\right),
\end{equation}
\begin{eqnarray}\label{eq:fieldmatter3}
&&8\pi p_t=\frac{\lambda}{6}\left(3\rho-p_r-8p_t\right)+\frac{1}{2r^2}\left(\frac{b}{r}-b'\right)+\\
&&+\frac{\zeta'}{4r}\left(2-\frac{b}{r}+b'\right)+\frac{\zeta'^2}{4}\left(1-\frac{b}{r}\right)+\frac{\zeta''}{2}\left(1-\frac{b}{2r}\right). \nonumber
\end{eqnarray}

Taking into consideration the explicit forms of the redshift function $\zeta\left(r\right)$ and the shape function $b\left(r\right)$ given in Eqs.\eqref{eq:def_zeta} and \eqref{eq:def_shape}, one verifies that at the throat $r=r_0$ the following boundary conditions apply
\begin{equation}\label{eq:NECt1}
\rho\left(r_0\right)+p_r\left(r_0\right)=-\frac{\beta+1}{r_0^2\left(\lambda+8\pi\right)},
\end{equation}
\begin{equation}\label{eq:NECt2}
\rho\left(r_0\right)+p_t\left(r_0\right)=-\frac{2\left(\beta-1\right)+\zeta_0\alpha\left(1+\beta\right)}{4r_0^2\left(\lambda+8\pi\right)}.
\end{equation}
Since $\beta$ is restricted to positive values, one verifies from Eq.\eqref{eq:NECt1} that in the GR case, i.e., $\lambda=0$, the combination $\rho+p_r$ is always negative at the throat, thus violating the NEC, see Eq.\eqref{eq:NEC}. However, in the general case $\lambda\neq 0$, one verifies that the first inequality in Eq.\eqref{eq:NEC}, along with Eq.\eqref{eq:NECt1}, impose a constraint on the parameter $\lambda$. Consequently, the second of Eq.\eqref{eq:NEC}, along with Eq.\eqref{eq:NECt2} then impose a constraint on the parameter $\zeta_0$. These two constraints take the forms
\begin{equation}\label{eq:rest1}
\lambda<-8\pi, \qquad\zeta_0>\frac{2\left(1-\beta\right)}{\alpha\left(\beta+1\right)}\equiv \zeta_c.
\end{equation}
If these two constraints are satisfied, then the wormhole solution considered satisfies the NEC at the throat, a feature unattainable in the GR limit. A particularly interesting consequence of the second constraint in Eq.\eqref{eq:rest1} is that in the limit $\alpha\to 0$, the bound $\zeta_c$ diverges to $+\infty$ if $\beta<1$, i.e., there is no value of $\zeta_0$ for which the NEC is satisfied at the throat. Furthermore, one can prove that in the limit $\left(\alpha,\beta\right)\to\left(0,1\right)$, the value of $\zeta_c$ is undefined, i.e., it depends on the path chosen to take the limit. Thus, if $\alpha=0$ one must restrict the analysis to $\beta>1$ . Depending on the combination of parameters considered, two different outcomes might arise: either the wormhole solution satisfies the NEC for the whole spacetime, i.e., for the whole range of the radial coordinate $r$; or the wormhole solution satisfies the NEC in a finite range of the radial coordinate around the throat, say $r<r_c$,  but violates it elsewhere, for $r>r_c$. In the latter case, a spacetime matching with an exterior vacuum spacetime is necessary to guarantee the physical relevance of the solution for the whole spacetime. This possibility is analyzed later in Sec.\ref{sec:matching}.

\subsection{Solutions satisfying the NEC everywhere}\label{sec:allNEC}

Since satisfying the NEC at the throat is not enough to guarantee the physical relevance of the solutions obtained, let us now look into the conditions necessary for the wormhole solution to satisfy the NEC for the whole spacetime. We will start by analyzing the combinations $\rho+p_r$ and $\rho+p_t$ independently and impose constraints on the parameters $\alpha$, $\beta$ and $\zeta_0$ that guarantee their positivity for the whole spacetime, and then we combine the results into a unified set of constraints. In the following, we will make use of a convenient redefinition of the radial coordinate as
\begin{equation}\label{eq:defx}
x=\frac{r_0}{r},
\end{equation}
which is confined in the range $x\in \left]0,1\right]$, where $x=1$ corresponds to the throat $r=r_0$, and $x\to0$ corresponds to the spacial infinity $r\to\infty$. The advantage of such a redefinition stands not only in the simplicity of the notation but also because it allows us to perform an analysis of the spacetime up to spacial infinity while keeping the value of the radial coordinate finite.

\subsubsection{Constraints from $\rho+p_r>0$}\label{sec:const1}

Let us start by analyzing the combination $\rho+p_r$. The condition $\rho+p_r>0$ subjected to the restriction $\lambda<-8\pi$, taking into account the forms of the matter quantities that can be extracted from Eqs. \eqref{eq:fieldmatter1} to \eqref{eq:fieldmatter3}, and the redefinition of the radial coordinate in Eq.\eqref{eq:defx}, can be written in the form
\begin{equation}\label{eq:cond1}
\alpha x^\alpha\zeta_0+x^{\beta+1}\left[1+\beta-\alpha x^\alpha\zeta_0\right]>0,
\end{equation}
This equation can be recast in the form of a constraint for the parameter $\zeta_0$, given a combination of $\alpha$, $\beta$ and $r_0$ in the form
\begin{equation}\label{eq:rest2}
\zeta_0>\frac{1+\beta}{\alpha}\ \max\left(\frac{x^{\beta-\alpha+1}}{x^{\beta+1}-1}\right)\equiv \zeta_{\text{min}}.
\end{equation}
If $\zeta_0>\zeta_\text{min}$, the combination $\rho+p_r$ does not have any zeroes, i.e., it does not change sign, for the whole range of the radial coordinate $r$. Combined with the fact that if $\zeta_0>\zeta_c$ from Eq.\eqref{eq:rest1}, which states that the combination $\rho+p_r$ is positive at the throat, if $\zeta_0>\max\left(\zeta_c,\zeta_\text{min}\right)$, then $\rho+p_r>0$ for the whole spacetime. For different combinations of $\alpha$ and $\beta$, it can happen that either $\zeta_c$ or $\zeta_\text{min}$ is the most restrictive bound on $\zeta_0$ (e.g., for $(\alpha,\beta)=(8,6)$ one has $\zeta_c>\zeta_\text{min}$, but for $(\alpha,\beta)=(6,8)$ one has $\zeta_\text{min}>\zeta_c$), and thus it is always necessary to verify both values independently.

\subsubsection{Constraints from $\rho+p_t>0$}\label{sec:const2}

Let us now look into the combination $\rho+p_t$. The condition $\rho+p_t>0$ subjected to the restriction $\lambda<-8\pi$, taking into account the forms of the matter variables extracted from Eqs. \eqref{eq:fieldmatter1} to \eqref{eq:fieldmatter3} and the redefinition of the radial coordinate in Eq.\eqref{eq:defx}, can be written as
\begin{eqnarray}\label{eq:cond2}
&&x^{\beta+1}\left[2\left(1-\beta\right)-\alpha\left(1+2\alpha+\beta\right)x^\alpha\zeta_0-\right.\\
&&\left.-\alpha^2x^{2\alpha}\zeta_0^2\right]+\alpha^2x^\alpha\zeta_0\left(2+x^\alpha\zeta_0\right)>0\nonumber
\end{eqnarray}
Similarly to Eq.\eqref{eq:cond1}, this equation can be recast in the form of a bound for $\zeta_0$. However, since in this case Eq.\eqref{eq:cond2} is quadratic in $\zeta_0$, this equation effectively imposes a double constraint on the value of $\zeta_0$. The parameter $\zeta_0$ is then constrained to be in the range
\begin{equation}\label{eq:rest3}
\zeta_-<\zeta_0<\zeta_+,   
\end{equation}
where the parameters $\zeta_\pm$ are defined as
\begin{equation}
\zeta_+=\min\left[g_+\left(x\right)\right],\qquad \zeta_-=\max\left[g_-\left(x\right)\right],
\end{equation}
where the functions $g_\pm\left(x\right)$ are given by
\begin{equation}\label{def_zetapm}
g_\pm\left(x\right) = \frac{1}{\alpha x^\alpha}\frac{B\left(x\right)\pm\sqrt{B\left(x\right)^2+A\left(x\right)C\left(x\right)}}{A\left(x\right)},
\end{equation}
and the functions $A$, $B$ and $C$ for a given combination of $\alpha$ and $\beta$ can be written in the forms
\begin{equation}
A\left(x\right)=2\left(1-x^{1+\beta}\right),
\end{equation}
\begin{equation}
B\left(x\right)=\left(1+2\alpha+\beta\right)x^{1+\beta}-2\alpha,
\end{equation}
\begin{equation}
C\left(x\right)=4\left(1-\beta\right)x^{1+\beta}.
\end{equation}
Similarly to the previous analysis, if $\zeta_0$ is restricted to the range $\zeta_-<\zeta_0<\zeta_+$, then the combination $\rho+p_t$ does not have any zeroes, i.e., it does not change sign. Furthermore, if $\zeta_0>\zeta_c$, which guarantees that $\rho+p_t>0$ at the throat, then the combination $\rho+p_t$ is positive for the whole spacetime. Interestingly, the function $g_-\left(x\right)$ increases monotonically in the interval $x\in\left]0,1\right]$, which implies that $\max\left[g_-\left(x\right)\right]=g_-\left(1\right)=\zeta_c$, and thus one effectively has $\zeta_-=\zeta_c$.

An interesting consequence of this analysis is that if $B^2+AC=0$ for some $x$, then there is a crossing $\zeta_+=\zeta_-$, Eq.\eqref{eq:rest3} becomes impossible to satisfy, and there is no value of $\zeta_0$ for which $\rho+p_t$ does not change sign. Since $B^2\geq 0$ and $A\geq 0$ in the range of parameters of interest, i.e., $x\in \left]0,1\right]$ and $\alpha,\beta>0$, these crossings can only occur if $C\leq 0$ or $B^2=AC=0$. Let us now analyze these possibilities.

\textit{(i)} Assume that the crossing occurs at $x=1$. In this case one has $A=0$ and the condition $B^2=0$ imposes a constraint on $\beta$ of the form $\beta=-1$. Since $\beta$ is constrained to be positive to preserve the asymptotic flatness of the solutions considered, this crossing is excluded from the analysis;

\textit{(ii)} Assume that the crossing occurs in the limit $x\to0$. In this case one has $C=0$ and the condition $B^2=0$ imposes a constraint on $\alpha$ of the form $\alpha=0$. Since $\alpha$ is also constrained to be positive to preserve the asymptotic flatness of the solutions considered, this crossing is excluded from the analysis;

\textit{(iii)} Assume that the crossing occurs for some $x\in\left]0,1\right[$. In this case one has $B^2\geq 0$ and $A>0$, and the crossing can only occur if $C\leq0$, i.e., if $\beta\leq1$. Indeed, if $\beta=1$ one has $C=0$ and the condition $B^2=0$ shows that the crossing occurs for $x=\sqrt{\alpha/\left(\alpha+1\right)}$. To perform a general analysis of the crossing, let us consider a coordinate transformation $u=x^{1+\beta}$. Since $\beta\geq 1$, the range of the coordinate $u$ is preserved to be $u\in \left]0,1\right]$. In terms of the coordinate $u$, the equation $B^2+AC=0$ can be written in the form
\begin{equation}
\left[2\alpha-\left(1+2\alpha+\beta\right)u\right]^2=8\left(u-1\right)\left(\beta-1\right)u.
\end{equation}
This is a quadratic equation for $u$ which features two roots $u_\pm$. In the range of parameters considered, i.e., $0<\beta<1$ and $\alpha>0$, one verifies that these roots are real and $u_\pm\in\left]0,1\right]$, the two roots degenerating into a single root in the particular case $\beta=1$ mentioned before. Consequently, one must restrict their analysis to $\beta>1$ to avoid these crossings and guarantee that there exists a range of values for $\zeta_0$ for which the combination $\rho+p_t$ does not change sign. Furthermore, since $\zeta_-=\zeta_c$, the condition that the combination $\rho+p_t$ is positive at the throat is automatically satisfied, and thus $\rho+p_t>0$ for the entire spacetime.

\subsubsection{Full set of constraints and solutions}

In the previous sections we have determined the necessary conditions to preserve the positivity of the combinations $\rho+p_r$ and $\rho+p_t$ for the whole spacetime. Let us now combine the results into a simplified set of constraints on the parameters $\alpha$, $\beta$ and $\zeta_0$ that allow one to find wormhole solutions satisfying the NEC for the whole spacetime. Note that throughout the analysis we have assumed that $\lambda<-8\pi$, a condition previously proven necessary to guarantee the validity of the NEC at the throat.

In Sec.\ref{sec:const1} we have verified that a necessary condition for $\rho+p_r>0$ is $\zeta_0>\max\left(\zeta_c,\zeta_\text{min}\right)$, whereas in Sec.\ref{sec:const2} we have obtained that a necessary condition for $\rho+p_t>0$ is $\zeta_c<\zeta_0<\zeta_+$. A necessary condition for these two constraints to be solvable simultaneously is that $\zeta_+>\zeta_\text{min}$. From Eqs.\eqref{eq:rest2} and \eqref{def_zetapm}, one verifies that $\zeta_\text{min}=\zeta_+=0$ in the parameter region $\alpha<\beta+1$. Since $\zeta_c<0$ in this parameter region, the only possible value of $\zeta_0$ allowing for solutions satisfying the NEC for the whole spacetime is $\zeta_0=0$, resulting in a trivial redshift function $\zeta\left(r\right)=0$. Although these are mathematically acceptable solutions, their physical relevance is limited, and thus we shall restrict our analysis to solutions with non-trivial redshift functions, i.e., we focus on the parameter region $\alpha\geq \beta+1$. In this parameter region, one verifies that the condition $\zeta_c>\zeta_\text{min}$ is always verified. Under these considerations, the set of constraints on the parameters $\alpha$, $\beta$ and $\zeta_0$ that allows for wormhole solutions satisfying the NEC for the whole spacetime becomes
\begin{equation}\label{eq:everywhere}
\zeta_c<\zeta_0<\zeta_+, \qquad \beta>1, \qquad \alpha\geq \beta+1, \qquad \lambda<-8\pi.
\end{equation}

The analysis conducted above can also be extended to include the verification of the WEC and SEC for the whole spacetime. Since these conditions also imply the NEC, an analysis of the full parameter space is not necessary and we can restrict the analysis to the parameter region already defined by Eq. \eqref{eq:everywhere}. In the following sections, we perform this analysis.

\subsection{Solutions satisfying the WEC everywhere}\label{sec:WEC}

Let us start by analyzing the WEC. For the WEC to be satisfied, the matter quantities must satisfy the conditions given in Eq.\eqref{eq:NEC} along with the extra restriction $\rho>0$. At the throat, the following boundary condition applies
\begin{equation}
 \rho(r_0)=-\frac{\lambda\alpha\zeta_0+\lambda\beta(16+\alpha\zeta_0)+96\pi\beta}{24r_0^2(\lambda+4\pi)(\lambda+8\pi)}.
\end{equation}
Under the constraints previously obtained for the NEC in Eq. \eqref{eq:everywhere}, one verifies that $\rho$ is always positive at the throat and no extra restriction is required for this purpose. One must now verify that the function $\rho$ also does not feature any zeroes to guarantee that the positiveness of $\rho$ remains for the entire spacetime.

The condition $\rho>0$, in combination with the previously imposed parameter bounds in Eq.\eqref{eq:everywhere}, and upon the redefinition of the radial coordinate in Eq.\eqref{eq:defx}, may be written explicitly as
\begin{equation}\label{eq:WEC}
\begin{split}
    &x^{1+\beta}\left[\alpha\left(1-2\alpha-\beta\right) x^\alpha\lambda\zeta_0-4\beta\left(24\pi+4\lambda\right)-\right.\\
    &\left.-\alpha^2x^{2\alpha}\lambda \zeta_0^2\right]+2\alpha\left(\alpha-1\right)x^\alpha\lambda\zeta_0+\alpha^2x^{2\alpha}\lambda\zeta_0^2>0.
\end{split}
\end{equation}
Similarly to the analysis of the NEC, one verifies that this equation is again quadratic in $\zeta_0$ and therefore imposes a double constraint on $\zeta_0$ of the form
\begin{equation}
\bar\zeta_-<\zeta_0<\bar\zeta_+,   
\end{equation}
where the parameters $\bar\zeta_\pm$ are defined as
\begin{equation}
\bar\zeta_+=\min\left[\bar g_+\left(x\right)\right],\qquad \bar\zeta_-=\max\left[\bar g_-\left(x\right)\right],
\end{equation}
where the functions $\bar g_\pm\left(x\right)$ are given by
\begin{equation}\label{def_WECzetapm}
\bar g_\pm\left(x\right) = \frac{1}{\alpha x^\alpha}\frac{\bar B\left(x\right)\pm\sqrt{\bar B\left(x\right)^2+\bar A\left(x\right)\bar C\left(x\right)}}{\bar A\left(x\right)},
\end{equation}
and the functions $\bar A$, $\bar B$ and $\bar C$ for a given combination of $\alpha$, $\beta$ and $\lambda$ are given by
\begin{equation}
\bar A\left(x\right)=-2\alpha\lambda\left(1-x^{1+\beta}\right),
\end{equation}
\begin{equation}
\bar B\left(x\right)=\alpha\lambda\left[2\left(\alpha-1\right)-\left(2\alpha+\beta-1\right)x^{1+\beta}\right],
\end{equation}
\begin{equation}
\bar C\left(x\right)=-32\alpha\beta x^{1+\beta}\left(6\pi+\lambda\right).
\end{equation}
Similarly to the previous NEC case, if $\zeta_0$ is restricted to the range $\bar\zeta_-<\zeta_0<\bar\zeta_+$, then the function $\rho$ does not have any zeroes. Given that we have previously proven that $\rho$ is positive at the origin in the parameter region of Eq.\eqref{eq:everywhere}, then one concludes that $\rho$ is positive for the whole spacetime, and the WEC is satisfied. Furthermore, one verifies that the function $\bar g_-\left(x\right)$ increases monotonically in the interval $x\in\left]0,1\right]$, and thus one has $\max\left[\bar g_-\left(x\right)\right]=\bar g_-\left(1\right)\equiv\bar\zeta_c$, where $\bar\zeta_c$ is given by
\begin{equation}
\bar\zeta_c=-\frac{16\beta\left(6\pi+\lambda\right)}{\alpha\lambda\left(1+\beta\right)}.
\end{equation}
Interestingly, one verifies that in the range of parameters of interest, see Eq.\eqref{eq:everywhere}, one has $\zeta_c>\bar\zeta_c$ and $\zeta_+<\bar\zeta_+$, which implies that the bounds on $\zeta_0$ arising from $\rho>0$ are weaker than the ones arising from the verification of the NEC. Thus, one concludes that if the parameters of the solution are chosen in a way as to satisfy the NEC for the whole spacetime, then $\rho$ will be positive everywhere and the WEC will also be satisfied for the whole spacetime. Note however that this is a one-directional implication and that the positivity of $\rho$ does not imply the verification of the NEC, e.g., one can find solutions with $\rho$ positive outside of the parameter region constrained by Eq. \eqref{eq:everywhere}. This compatibility between the NEC and the condition $\rho>0$ also guarantees that no crossings of the form $\bar\zeta_+=\bar\zeta_-$ occur.

\subsection{Solutions satisfying the SEC everywhere}\label{sec:SEC}

Let us turn now to the SEC. For the SEC to be satisfied, the matter quantities must satisfy the conditions given in Eq.\eqref{eq:NEC}, with the extra restriction $\rho+p_r+2p_t>0$. This condition computed at the throat gives rise to the following boundary condition
\begin{equation}\label{eq:SECthroat}
\begin{split}
&\rho\left(r_0\right)+p_r\left(r_0\right)+2p_t\left(r_0\right)=\\
&=-\frac{8\beta\lambda+\alpha\left(1+\beta\right)\zeta_0\left(24\pi+5\lambda\right)}{12r_0^2\left(\lambda+8\pi\right)\left(\lambda+4\pi\right)}.
\end{split}
\end{equation}
Unlike for the WEC, the conditions previously obtained in Eq.\eqref{eq:everywhere} are not sufficient to guarantee that Eq.\eqref{eq:SECthroat} is positive at the throat. Indeed, one verifies that if $\beta>5$ then $\lambda$ must satisfy one extra constraint given by
\begin{equation}
    \lambda>-24\pi\frac{\alpha\left(1+\beta\right)\zeta_0}{5\alpha\zeta_0\left(1+\beta\right)+8\beta}\equiv \lambda_{\min},
\end{equation}
which depends on the value of $\zeta_0$. In particular, taking the lowest bound $\zeta_0=\zeta_c$, this constraint takes the form
\begin{equation}\label{eq:boundlambda1}
\lambda>24\pi\frac{1-\beta}{\beta-5}\equiv \lambda_0,
\end{equation}
which guarantees that Eq.\eqref{eq:SECthroat} is positive independently of the value of $\zeta_0$. In combination with the first of Eq.\eqref{eq:rest1}, this constraint can be rewritten as a bound on $\lambda$ of the form $\lambda_0<\lambda<-8\pi$, which is a more restrictive bound than the previously found $\lambda<-8\pi$. Note also that $\lambda_0$ increases monotonically with $\beta$ for $\beta>5$, achieving a maximum value of $\lambda_0\left(\beta\to\infty\right)=-24\pi$, which implies that independently of the value of $\beta$ it is always possible to find a suitable value of $\lambda$ in the range $\lambda_0<\lambda<-8\pi$. On the other hand, if $\beta\leq 5$, no extra requirements are necessary. If these requirements are met, the function $\rho+p_r+2p_t$ is positive at the throat. One must now verify which conditions are necessary for this function to not have any zeroes, to extrapolate that the condition holds for the entire spacetime.

The condition $\rho+p_r+2p_t>0$, along with the parameter bounds obtained in Eq.\eqref{eq:everywhere} and upon a redefinition of the radial coordinate as in Eq.\eqref{eq:defx}, takes the form
\begin{equation}
    \begin{split}
        &-\alpha x^\alpha\left[2+\left(\beta-1\right)x^{1+\beta}\right]\left(24\pi+5\lambda\right)\zeta_0-8\alpha \lambda x^{1+\beta}+\\
&+\alpha^2x^\alpha\left(1-x^{1+\beta}\right)\left(2+x^\alpha\zeta_0\right)\left(24\pi+5\lambda\right)\zeta_0>0.
    \end{split}
\end{equation}
Analogously to the previous cases analyzed, this is a quadratic equation and imposes a double constraint on $\zeta_0$ of the form,
\begin{equation}
\hat\zeta_-<\zeta_0<\hat\zeta_+,   
\end{equation}
where the parameters $\hat\zeta_\pm$ are defined as
\begin{equation}
\hat\zeta_+=\min\left[\hat g_+\left(x\right)\right],\qquad \hat\zeta_-=\max\left[\hat g_-\left(x\right)\right],
\end{equation}
where the functions $\hat g_\pm\left(x\right)$ are given by
\begin{equation}\label{def_SECzetapm}
\hat g_\pm\left(x\right) = \frac{1}{\alpha x^\alpha}\frac{\hat B\left(x\right)\pm\sqrt{\hat B\left(x\right)^2+\hat A\left(x\right)\hat C\left(x\right)}}{\hat A\left(x\right)},
\end{equation}
and the functions $\hat A$, $\hat B$ and $\hat C$ for a certain combination of $\alpha$, $\beta$ and $\lambda$ are now written as
\begin{equation}
\hat A\left(x\right)=-2\alpha\left(1-x^{1+\beta}\right)\left(24\pi+5\lambda\right),
\end{equation}
\begin{equation}
\hat B\left(x\right)=\alpha\left[2\left(\alpha-1\right)-\left(2\alpha+\beta-1\right)x^{1+\beta}\right]\left(24\pi+5\lambda\right),
\end{equation}
\begin{equation}
\hat C\left(x\right)=-16\alpha\beta\lambda x^{1+\beta}.
\end{equation}
Similarly to the previous analyses, if $\zeta_0$ is restricted to the range $\hat\zeta_-<\zeta_0<\hat\zeta_+$, then the function $\rho+p_r+2p_t$ does not have any zeroes, which in combination with the restrictions in Eq.\eqref{eq:everywhere} and, if $\beta >5$, also Eq.\eqref{eq:SECthroat}, which guarantee the positivity at the throat, imply that the SEC is satisfied for the whole spacetime. Also, since the function $\hat g_-\left(x\right)$ increases monotonically in the interval $x\in\left]0,1\right]$, then again one verifies that $\max\left[\hat g_-\left(x\right)\right]=\hat g_-\left(1\right)\equiv\hat\zeta_c$, where $\hat\zeta_c$ is defined as
\begin{equation}
\hat\zeta_c=-\frac{8\beta\lambda}{\alpha\left(1+\beta\right)\left(24\pi + 5 \lambda\right)}.
\end{equation}
For $\beta<5$, one verifies that in the range of parameters of interest, see Eq.\eqref{eq:everywhere}, one has $\zeta_c>\hat\zeta_c$, and thus the lower bound on $\zeta_0$ arising from $\rho+p_r+2p_t>0$ is weaker than the one arising from the NEC. If $\beta=5$, one verifies that in the limit $\lambda\to -\infty$ the bounds become equal, i.e., $\zeta_c=\hat\zeta_c$, and if $\beta>5$ there is a value $\lambda_c^-$ for which if $\lambda<\lambda_c^-$, then one has $\zeta_c<\hat\zeta_c$, thus implying that the lower bound on $\zeta_0$ arising from $\rho+p_r+2p_t>0$ is stronger than the one arising from the NEC. The value of $\lambda_c^-$ depends solely on $\beta$ in the form
\begin{equation}
    \lambda_c^-=24\pi\frac{1-\beta}{\beta-5}=\lambda_0.
\end{equation}
Thus, for the cases where $\beta>5$, one concludes that if $\lambda$ satisfies the condition $\lambda_{\min}<\lambda<\lambda_0$, then the lower bound on $\zeta_0$ arising from $\rho+p_r+2p_t>0$ is stronger than the one arising from the NEC, whereas if $\lambda$ satisfies the condition $\lambda_0<\lambda<-8\pi$ it is the bound arising from the NEC that is stronger. Since from Eq.\eqref{eq:boundlambda1} we have already concluded that the condition $\rho+p_r+2p_t>0$ is only satisfied at the origin independently of the value of $\zeta_0$ if $\lambda>\lambda_0$, then by restricting the analysis to this range we guarantee that the lower bound on $\zeta_0$ arising from the NEC is always stronger.

A similar situation arises for the upper bound $\hat\zeta_+$. For the same range of parameters of interest, if $\beta\leq5$ one verifies that $\zeta_+<\hat\zeta_+$ and thus the upper bound on $\zeta_0$ arising from the NEC is stronger than the one arising from $\rho+p_r+2p_t>0$. However, if $\beta>5$, one finds that for a given combination of $\alpha$ and $\beta$ there is another critical value $\lambda_c^+$ for which if $\lambda<\lambda_c^+$ one has $\hat\zeta_+<\zeta_+$, implying that the upper bound on $\zeta_0$ arising from $\rho+p_r+2p_t>0$ is stronger than the one arising from the NEC. The value of $\lambda_c^+$ is given by
\begin{equation}\label{eq:lambdacp}
    \lambda_c^+=24\pi\frac{\left(\alpha-1\right)\left(\beta-1\right)}{5\left(\beta-1\right)-\alpha\left(\beta-5\right)}.
\end{equation}
Note that not all combinations of parameters with $\beta>5$ will give rise to a $\lambda_c^+<-8\pi$, i.e., the critical value $\lambda_c^+$ might fall outside of the range of parameters of interest. Indeed, from Eq.\eqref{eq:lambdacp}, one verifies that in order to obtain a $\lambda_c^+$ in the range $\lambda_c^+<-8\pi$ for a fixed $\beta>5$, one needs $\alpha>\alpha_c$, where $\alpha_c$ is given in terms of $\beta$ as
\begin{equation}\label{eq:alphac}
    \alpha_c=5\frac{\beta-1}{\beta-5}.
\end{equation}
Now, let us analyze the consequences of these results. For a fixed $\beta>5$ and $\alpha>\alpha_c$, which guarantees that $\lambda_c^+<-8\pi$, one verifies that $\lambda_c^+$ from Eq. \eqref{eq:lambdacp} is a monotonically increasing function of $\alpha$, achieving a maximum $\lambda_c^+\left(\alpha\to\infty\right)=\lambda_0$. Let us define an interval $I$ as $I=\left]\lambda_{\min},\lambda_c^+\right[$. Depending on the values of $\alpha$, $\beta$, and $\zeta_0$, the interval $I$ might be either empty or finite. If $I$ is finite for a given combination of parameters $\alpha$, $\beta$, and $\zeta_0$, then for $\lambda\in I$ the upper bound on $\zeta_0$ arising from $\rho+p_r+2p_t>0$ is stronger than the one arising from the NEC. If the interval $I$ is empty, then for any $\lambda>\lambda_{\min}$ the upper bound on $\zeta_0$ arising from the NEC will be stronger. Again, by restricting our analysis to the region $\lambda_0<\lambda<-8\pi$, then one guarantees that independently of the values of $\beta>5$ and $\alpha>\alpha_c$ it is impossible to find a value of $\lambda$ that satisfies $\lambda>\lambda_0$ and $\lambda<\lambda_c^+$ simultaneously. Consequently, in this region there are no possible combinations of parameters $\alpha$ and $\beta$ within the range of parameters of interest for which the upper bound on $\zeta_0$ arising from $\rho+p_r+2p_t>0$ is stronger than the one arising from the NEC. This conclusion, combined with the one obtained in the previous paragraph for the lower bound on $\zeta_0$, implies that if the parameters of the solution are chosen in a way as to satisfy the NEC for the whole spacetime and $\lambda$ is chosen in the region $\lambda_0<\lambda<8\pi$, then $\rho+p_r+2p_t$ will be positive everywhere independently of the values of $\zeta_0$, $\alpha$ and $\beta$, and the SEC will also be satisfied for the whole spacetime. Again, note that this is a one-directional implication, and thus finding a solution for which $\rho+p_r+2p_t>0$ does not guarantee that the NEC is satisfied. This compatibility between the NEC and the condition $\rho+p_r+2p_t>0$ also guarantees that no crossings $\hat\zeta_+=\hat\zeta_-$ occur.

\subsection{Solutions satisfying the DEC everywhere}\label{sec:DEC}

Let us finally consider the DEC. For the DEC to be satisfied, the matter quantities must satisfy the condition in Eq.\eqref{eq:NEC}, as well as the extra restrictions $\rho>|p_r|$ and $\rho>|p_t|$. Note that these two conditions also imply that $\rho>0$, and thus the DEC not only implies the NEC but also the WEC. Now, due to the complicated dependence of $p_r$ and $p_t$ on the parameters of the model, the signs of $p_r$ and $p_t$ are difficult to determine without explicitly inputting the values of the parameters. Thus, solving $\rho>|p_r|$ and $\rho>|p_t$ in general can prove to be a difficult task. However, note that if any of the pressures $p_i$ is negative at a given point, then $\rho+p_i>0$ immediately implies that $\rho>|p_i|$ at that point. On the other hand, if any of the pressures $p_i$ is positive at some point, then proving that $\rho-p_i>0$ would be sufficient to state that $\rho>|p_i|$ at that point. Indeed, the DEC states that there is a dominance of the energy density $\rho$ over the pressures $p_r$ and $p_t$ and thus proving separately that $\rho+p_i>0$ and $\rho-p_i>0$ for the whole spacetime automatically proves that $\rho>|p_i|$ for the whole spacetime. In the previous sections, we have already determined the necessary conditions for $\rho+p_r>0$ and $\rho+p_t>0$, and thus in this section we can restrict our analysis to the study of the two conditions $\rho-p_r>0$ and $\rho-p_t>0$.

Following the same method as before, let us first analyze the positivity at the throat. At the throat, we thus obtain the following boundary conditions
\begin{equation}\label{eq:deccond1}
\begin{split}
    &\rho\left(r_0\right)-p_r\left(r_0\right)=\\
    &=-\frac{48\pi\left(\beta-1\right)+\lambda\left[\alpha\zeta_0\left(1+\beta\right)+4\beta-12\right]}{12r_0^2\left(\lambda+8\pi\right)\left(\lambda+4\pi\right)},
\end{split}
\end{equation}
\begin{equation}\label{eq:deccond2}
\begin{split}
    &\rho\left(r_0\right)-p_t\left(r_0\right)=\\
    &=\frac{6\pi\left[\alpha\zeta_0\left(1+\beta\right)-6\beta-2\right]+\lambda\left[\alpha\zeta_0\left(1+\beta\right)-5\beta-3\right]}{6r_0^2\left(\lambda+8\pi\right)\left(\lambda+4\pi\right)}.
\end{split}
\end{equation}
In the range of parameters of interest, see Eq.\eqref{eq:everywhere}, one verifies that the condition in Eq.\eqref{eq:deccond2} is always positive. Indeed, one would need $\lambda>-8\pi$ in order to obtain $\rho-p_t<0$ at the throat. Thus, this boundary condition does not impose any extra restrictions on the parameter space. On the other hand, the same is not true for Eq.\eqref{eq:deccond1}. For this condition to be positive, $\zeta_0$ is required to satisfy the constraint
\begin{equation}\label{eq:DECthroat}
\zeta_0>\frac{48\pi\left(1-\beta\right)+4\lambda\left(3-\beta\right)}{\alpha\lambda\left(1+\beta\right)}\equiv\tilde\zeta_c,
\end{equation}
which depends explicitly on the value of $\lambda$. If this constraint is satisfied, along with the constraints given in Eq.\eqref{eq:everywhere}, one guarantees that the DEC is satisfied at the throat. Similarly to the previous cases, one must now verify what are the constraints on the parameters $\alpha$, $\beta$, $\lambda$ and $\zeta_0$ that guarantee that the combinations $\rho-p_r$ and $\rho-p_t$ do not have any zeroes and, consequently, remain positive for the whole spacetime.

\subsubsection{Constraints from $\rho>|p_r|$}

Let us start by analyzing the condition $\rho-p_r>0$, which associated with the previously studied $\rho+p_r>0$ implies that $\rho>|p_r|$. The condition $\rho-p_r>0$, along with the parameter bounds obtained in Eq.\eqref{eq:everywhere} and upon a redefinition of the radial coordinate as in Eq.\eqref{eq:defx}, takes the form
\begin{equation}
\begin{split}
&\alpha x^\alpha\zeta_0\left\{\left[48\pi+\alpha\lambda\left(2+x^\alpha\zeta_0\right)\right]\left(1-x^{1+\beta}\right)+\lambda\left[10-\right.\right.\\
&\left.\left.-\left(11+\beta\right)x^{1+\beta}\right]\right\}
-4x^{1+\beta}\left[\left(12\pi+\lambda\right)\left(\beta-1  \right)-2\lambda\right]>0.
\end{split}
\end{equation}
Again, this equation is quadratic in $\zeta_0$ and thus imposes a double constraint on $\zeta_0$ of the form
\begin{equation}
\tilde\zeta_- <\zeta_0<\tilde\zeta_+ ,
\end{equation}
where the parameters $\tilde\zeta_\pm $ are defined as
\begin{equation}
\tilde\zeta_+ =\min\left[\tilde g_+ \left(x\right)\right],\qquad \tilde\zeta_- =\max\left[\tilde g_- \left(x\right)\right],
\end{equation}
where the functions $\tilde g_\pm \left(x\right)$ are given by
\begin{equation}\label{def_SECzetapm}
\tilde g_\pm \left(x\right) = \frac{1}{\alpha x^\alpha}\frac{\tilde B \left(x\right)\pm\sqrt{\tilde B \left(x\right)^2+\tilde A \left(x\right)\tilde C \left(x\right)}}{\tilde A \left(x\right)},
\end{equation}
and the functions $\tilde A $, $\tilde B $ and $\tilde C $ for a certain combination of $\alpha$, $\beta$ and $\lambda$ are now written as
\begin{equation}
\tilde A \left(x\right)=-2\lambda\left(1-x^{1+\beta}\right),
\end{equation}
\begin{equation}
\tilde B \left(x\right)=48\pi+2\lambda\left(\alpha+5\right)-x^{1+\beta}\left[48\pi+\lambda\left(11+2\alpha+\beta\right)\right],
\end{equation}
\begin{equation}
\tilde C \left(x\right)=-8x^{1+\beta}\left[\left(12\pi+\lambda\right)\left(\beta-1\right)-2\lambda\right].
\end{equation}
Again, if $\zeta_0$ is restricted to the range $\tilde\zeta_- <\zeta_0<\tilde\zeta_+ $, then the function $\rho-p_r$ does not have any zeroes, which in combination with the conditions in Eq.\eqref{eq:everywhere} and $\zeta>\tilde\zeta_c$ that guarantee the positivity at the throat implies that $\rho>|p_r|$ for the whole spacetime. The function $\tilde g_- $ also increases monotonically in the interval $x\in\left]0,1\right]$ in the range of parameters of interest, and thus $\max\left[\tilde g_- \left(x\right)\right]=\tilde g_- \left(1\right)=\tilde \zeta_c$, where $\tilde\zeta_c$ was previously defined in Eq.\eqref{eq:DECthroat}.

Similarly to what happened in Sec.\ref{sec:const2}, in this case it is necessary to verify if at any point $x$ one obtains $(\tilde B )^2+\tilde A \tilde C =0$, as these correspond to the crossings $\tilde\zeta_+ =\tilde\zeta_- $ and prevent one from finding a suitable value of $\zeta_0$. The dependence of the functions $\tilde A $, $\tilde B $ and $\tilde C $ on $\lambda$ imply that the signs of these functions are not determined in the range of parameters of interest, and thus it is necessary to perform this analysis in general. Taking a coordinate transformation of the form $u=x^{1+\beta}$, the equation $(\tilde B )^2+\tilde A \tilde C =0$ can be rewritten in the form
\begin{equation}
\begin{split}
&\left[48\pi\left(u-1\right)-2\lambda\left(5+\alpha\right)+u \lambda\left(11+2\alpha+\beta\right)\right]^2=\\
&=16u\lambda\left(u-1\right)\left[12\pi\left(\beta-1\right)-\lambda\left(\beta-3\right)\right].
\end{split}
\end{equation}
This is a quadratic equation for $u$ that features two roots $\tilde u_\pm$. One verifies that for any $\beta\leq 3$, the two roots $\tilde u_\pm$ are real and $\tilde u_\pm\in\left]0,1\right]$, independently of the values of $\alpha$ and $\lambda$, thus implying that there will be crossings $\tilde\zeta_+ =\tilde\zeta_- $, and consequently the function $\rho-p_r$ will change sign at some finite radius $r$. To avoid these crossings, one must thus restrict the analysis to the region $\beta>3$. However, this restriction is not enough. Even if one chooses some combination of $\alpha$ and $\beta$ such $\beta>3$ and $\alpha\geq\beta+1$, the crossings the roots $\tilde u_\pm$ might still be real and in the interval $\tilde u_\pm\in\left]0,1\right]$. One then verifies that in order to avoid these roots, one must impose a constraint on $\lambda$ of the form
\begin{equation}\label{eq:DECboundlambda}
\lambda < 12\pi\frac{1-\beta}{\beta-3}\equiv\tilde\lambda_0.
\end{equation}
If this condition is satisfied, then one guarantees that no crossings $\tilde\zeta_+ =\tilde\zeta_- $ occur and it is always possible to choose an appropriate value of $\zeta_0$ such that the condition $\rho-p_r$ does not have any zeroes. Furthermore, since $\tilde\zeta_- =\tilde\zeta_c$, one guarantees that $\rho-p_r>0$ at the throat, and thus $\rho>|p_r|$ for the whole spacetime. Under these considerations, the set of constraints on the parameters $\alpha$, $\beta$, $\zeta_0$ and $\lambda$ necessary for $\rho>|p_r|$ and the NEC to be satisfied for the whole spacetime are
\begin{equation}\label{eq:DECeverywhere}
\tilde\zeta_c<\zeta_0<\tilde\zeta_+,\qquad \beta>3, \qquad \alpha\geq\beta+1,\qquad\lambda<\tilde\lambda_0.
\end{equation}
Note that by definition the DEC implies the WEC and the NEC, but it does not imply the SEC, meaning that if one is looking for a solution satisfying all four energy conditions, it is necessary to combine these results with the ones from Sec.\ref{sec:SEC}.

\subsubsection{Constraints from $\rho>|p_t|$}

Let us repeat the analysis for the condition $\rho-p_t>0$, which in combination with the condition $\rho+p_t>0$ analyzed before implies that $\rho>|p_t|$. This condition, within the parameter bounds obtained in Eq.\eqref{eq:everywhere} and upon a redefinition of the radial coordinate as in Eq.\eqref{eq:defx}, takes the form 
\begin{equation}\label{eq:DECpt}
\begin{split}
&\alpha x^{\alpha}\zeta_0\left\{x^{1+\beta}\left(6\pi+\lambda\right)\left(1+\beta\right)-\left[\lambda+\alpha\left(2+x^\alpha\zeta_0\right)\left(6\pi+\lambda\right)\right]\right.\\
&\left.\times\left(1-x^{1+\beta}\right)\right\}-x^{1+\beta}\left[\left(12\pi+\lambda\right)\left(1+3\beta\right)+2\lambda\left(1+\beta\right)\right]>0.
\end{split}
\end{equation}
Similarly to the previous sections, this equation is quadratic in $\zeta_0$ and could be used to impose bounds on the value of this parameter, following the same procedure as before. However, such an analysis is not necessary, as one verifies that, in the range of parameters of interest, i.e. $\zeta_0>\tilde\zeta_c$, $\beta>3$, $\alpha\geq\beta+1$, and $\lambda<\tilde\lambda_0$, Eq.\eqref{eq:DECpt} does not feature any zeroes in the interval $x\in\left]0,1\right]$. Indeed, the second term in the equation is always positive and relatively large due to its proportionality to $-\lambda$ and $\beta$, whereas the first term, even though it can be either positive or negative, is bounded to smaller absolute values due to the proportionality in $\alpha\zeta_0$. Consequently, if the matter quantities satisfy the NEC and the condition $\rho>|p_r|$ for the whole spacetime, then the condition $\rho>|p_t|$ is automatically satisfied for the whole spacetime. One may indeed analytically find conditions for the existence of zeroes (or lack thereof) in Eq.\eqref{eq:DECpt}. This produces a set of constraints on the parameters $\zeta_0$, $\beta$ (or $\alpha$), and $\lambda$, which we do not show explicitly due to their size. 

\subsection{Explicit examples of solutions}

In the previous sections we have derived the necessary conditions for a wormhole solution to satisfy the NEC, WEC, SEC, and DEC for the whole spacetime. Surprisingly, we have verified that by restricting our analysis to the range of $\lambda$ given by $\lambda_0<\lambda<-8\pi$, then the satisfaction of the NEC for the whole spacetime automatically guarantees that both the WEC and the SEC are also satisfied. Furthermore, one verifies that $\hat\zeta_+<\bar\zeta_+$ and that $\hat\zeta_->\bar\zeta_-$ in the range of parameters of interest, see Eq.\eqref{eq:everywhere}, which implies that the bounds on $\zeta_0$ arising from the SEC are stronger than the ones arising from the WEC. The same is not true for the DEC, as we have verified that the latter requires a stronger bound on the parameters $\beta$, $\zeta_0$, and $\lambda$. A general recipe to obtain solutions satisfying the desired energy conditions is the following:

\textit{Solution satisfying NEC and WEC:}
\begin{enumerate}
    \item Choose $\beta>1$ and $\lambda<-8\pi$;
    \item Choose $\alpha>\beta+1$;
    \item Choose $\zeta_c<\zeta_0<\zeta_+$.
\end{enumerate}

\textit{Solution satisfying NEC, WEC, and SEC:}
\begin{enumerate}
    \item Choose $\beta>1$;
    \item Choose $\alpha>\beta+1$;
    \item If $\beta\leq 5$, choose  $\lambda<-8\pi$;
    \item If $\beta>5$, choose $\lambda_0<\lambda<-8\pi$;
    \item Choose $\zeta_c<\zeta_0<\zeta_+$.
\end{enumerate}

\textit{Solution satisfying NEC, WEC, SEC, and DEC:}
\begin{enumerate}
    \item Choose $\beta>3$;
    \item Choose $\alpha>\beta+1$;
    \item If $\beta\leq 5$, choose  $\lambda<-\tilde\lambda_0$;
    \item If $\beta>5$, choose $\lambda_0<\lambda<\tilde\lambda_0$;
    \item Choose $\tilde\zeta_c<\zeta_0<\tilde\zeta_+$.
\end{enumerate}
In the following, we will provide two examples of solutions: one satisfying the NEC with $\beta=3$, which consequently satisfies the WEC and the SEC, but not the DEC; and another solution also satisfying the DEC.

\begin{figure*}[h]
    \centering
    \includegraphics[scale=0.4]{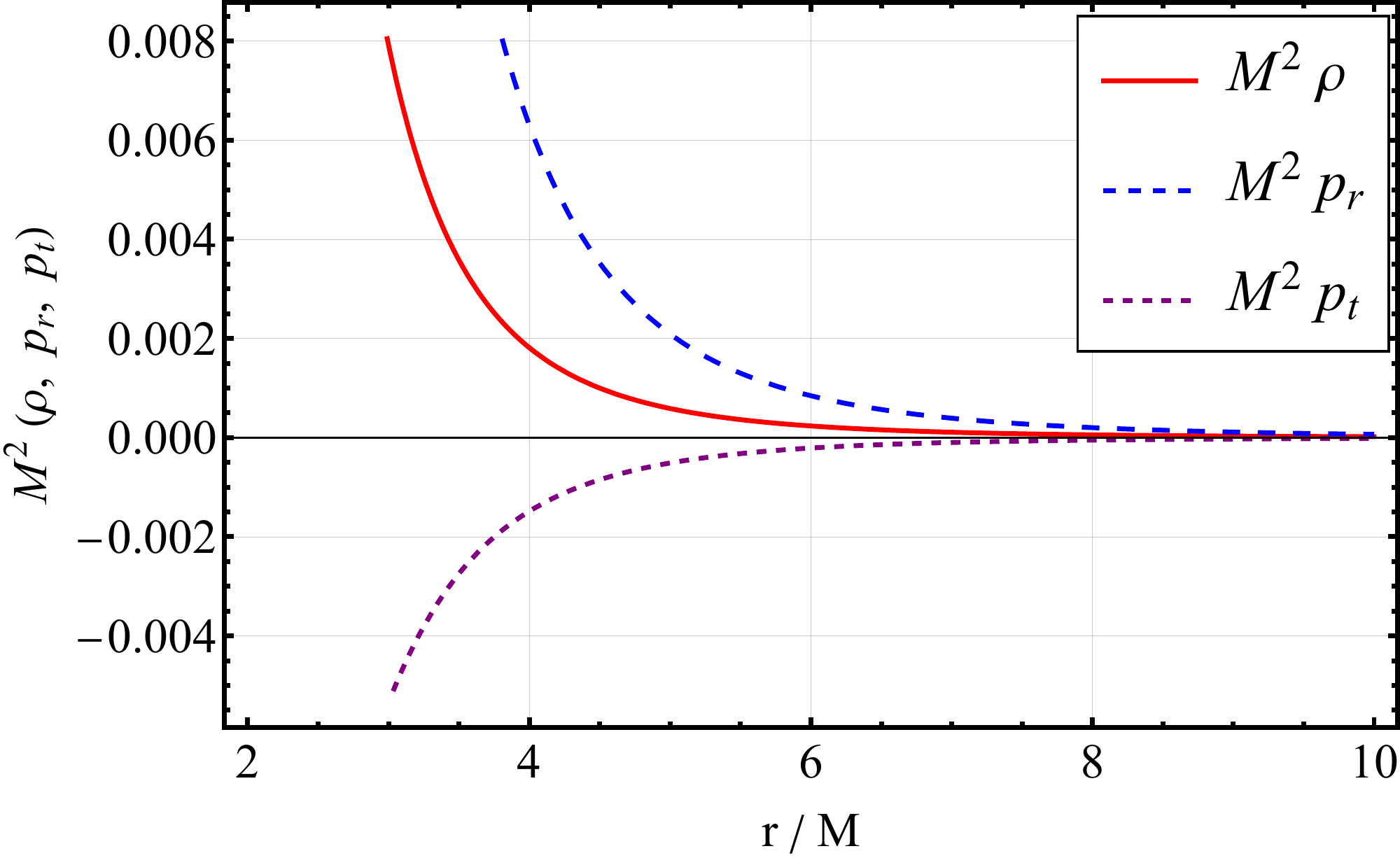}
    \includegraphics[scale=0.4]{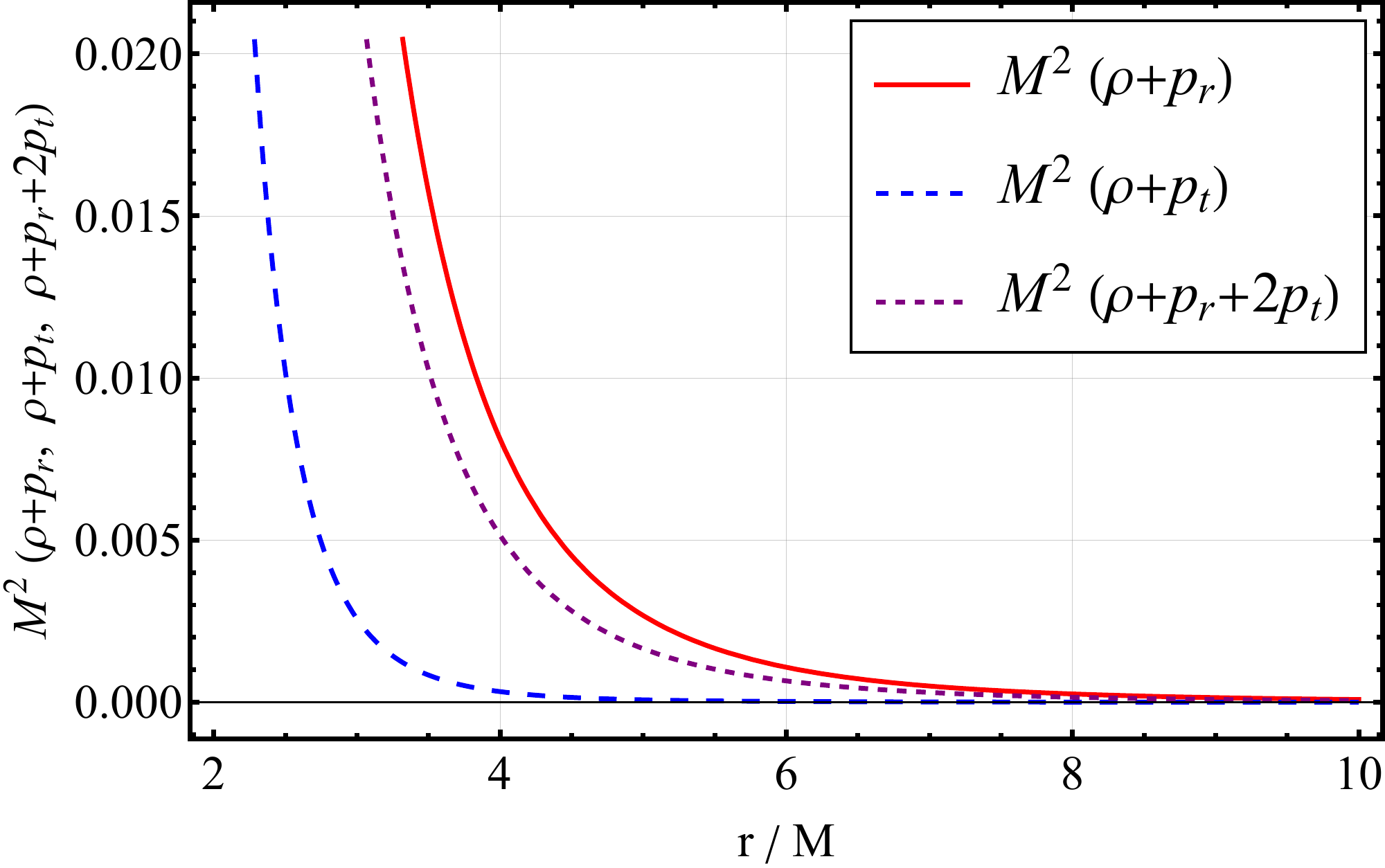}
    \caption{Matter components $\rho$, $p_r$ and $p_t$ (left panel) and combinations $\rho+p_r$, $\rho+p_t$, and $\rho+p_r+2p_t$ (right panel) as functions of the normalized radial coordinate $r/M$ with $\alpha=3$, $\beta=2$, $\lambda=-9\pi$, $r_0=2M$, and $\zeta_0=0.1$. Since $\rho+p_r>0$ and $\rho+p_t>0$, one verifies that the NEC is satisfied. Furthermore, since $\rho>0$ and $\rho+p_r+2p_t>0$, the WEC and SEC are also satisfied for the whole spacetime.}
    \label{fig:solution1}
\end{figure*}

\textit{Solution 1: } According to the restrictions obtained in Eq.\eqref{eq:everywhere} for $\alpha$, $\beta$, and $\lambda$, let us consider $\alpha=3$, $\beta=2$, and $\lambda=-9\pi$. Furthermore, let us take $r_0=2M$, for some constant $M$ with units of mass. For this combination of parameters, we obtain $\zeta_c\sim -0.22$ and $\zeta_+\sim 0.11$. Thus, the wormhole will satisfy the NEC if $-0.22<\zeta_0<0.11$. Regarding the WEC, one verifies that $\bar\zeta_-\sim -1.19$ and $\bar\zeta_+\sim 0.88$. These two bounds are weaker than the bounds imposed by the NEC, as expected from the results of Sec.\ref{sec:WEC}. For the SEC one verifies that $\hat\zeta_-\sim-0.76$ and $\hat\zeta_+\sim 0.39$. Again, these bounds are weaker than the ones arising from the NEC, as expected according to the results of Sec. \ref{sec:SEC}. One can thus choose $\zeta_0=0.1$ to complete the solution. The solutions for the matter variables $\rho$, $p_r$ and $p_t$, as well as the combinations $\rho+p_r$, $\rho+p_t$, and $\rho+p_r+2p_t$ necessary for the NEC, WEC and SEC, are plotted in Fig.\ref{fig:solution1}. One can verify that the NEC, WEC, and SEC are satisfied for the entire range of the radial coordinate without the necessity of performing a matching with an exterior vacuum spacetime, but that the DEC is violated since $\rho<p_r$.

\begin{figure*}[h]
    \centering
    \includegraphics[scale=0.4]{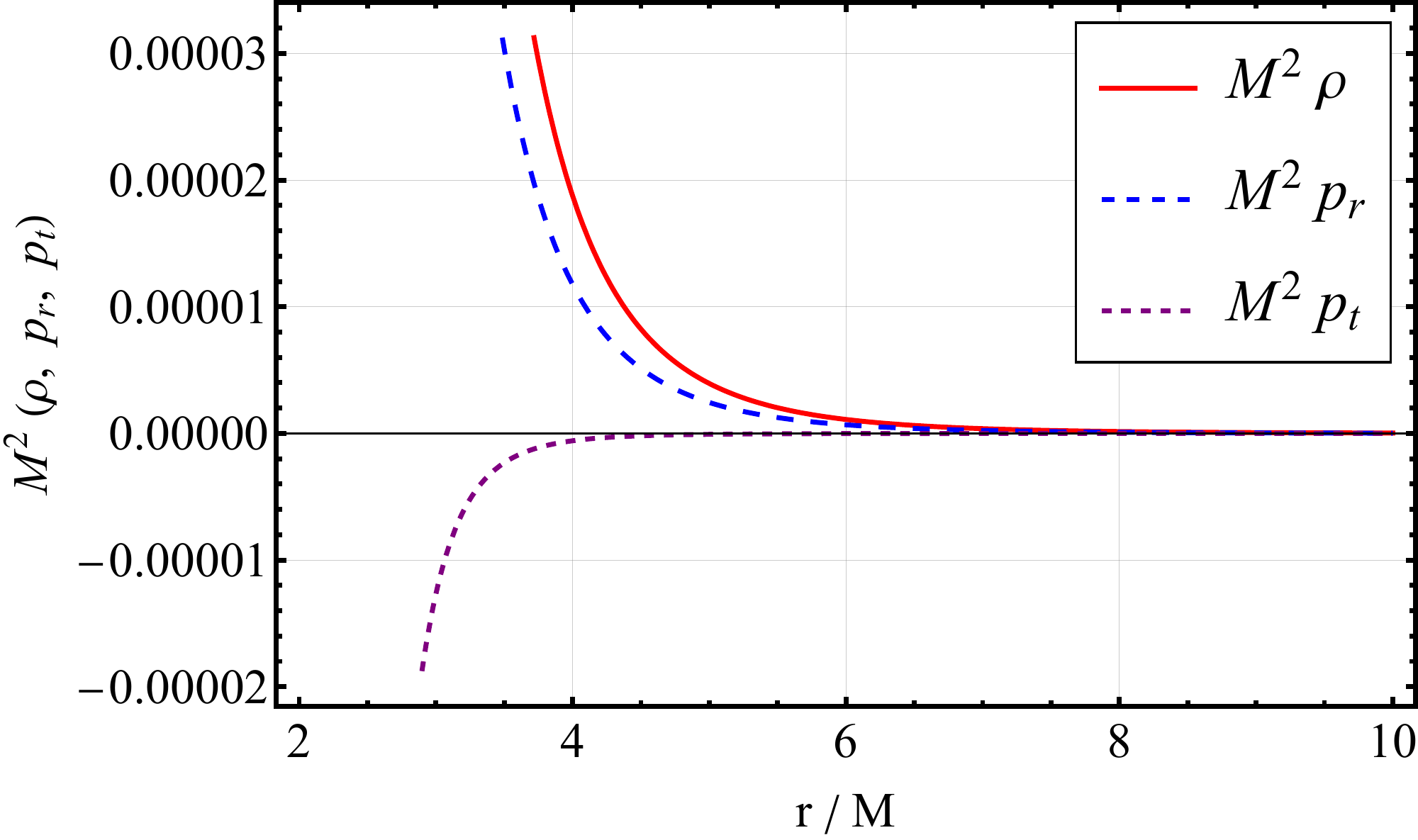}
    \includegraphics[scale=0.4]{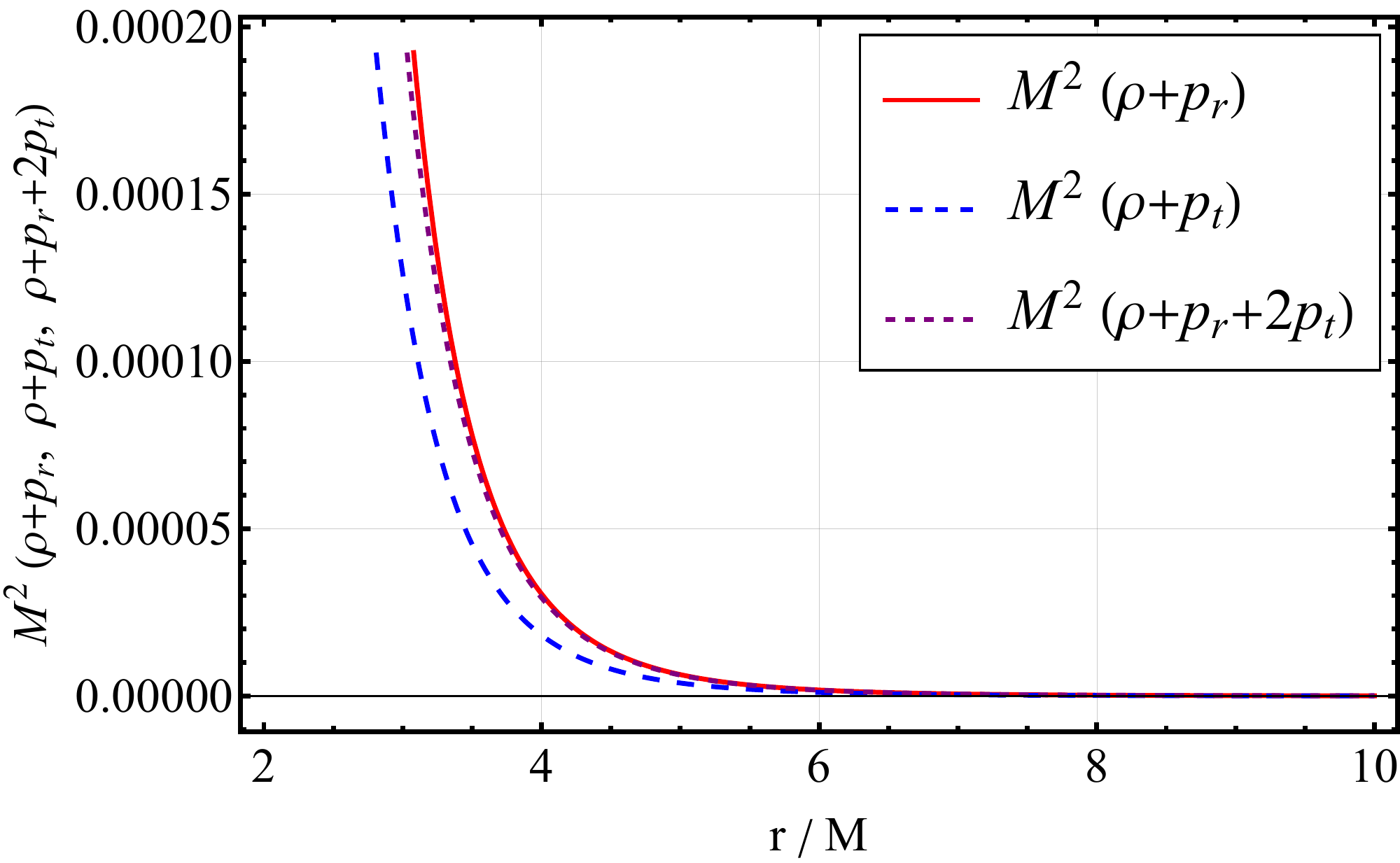}
    \includegraphics[scale=0.4]{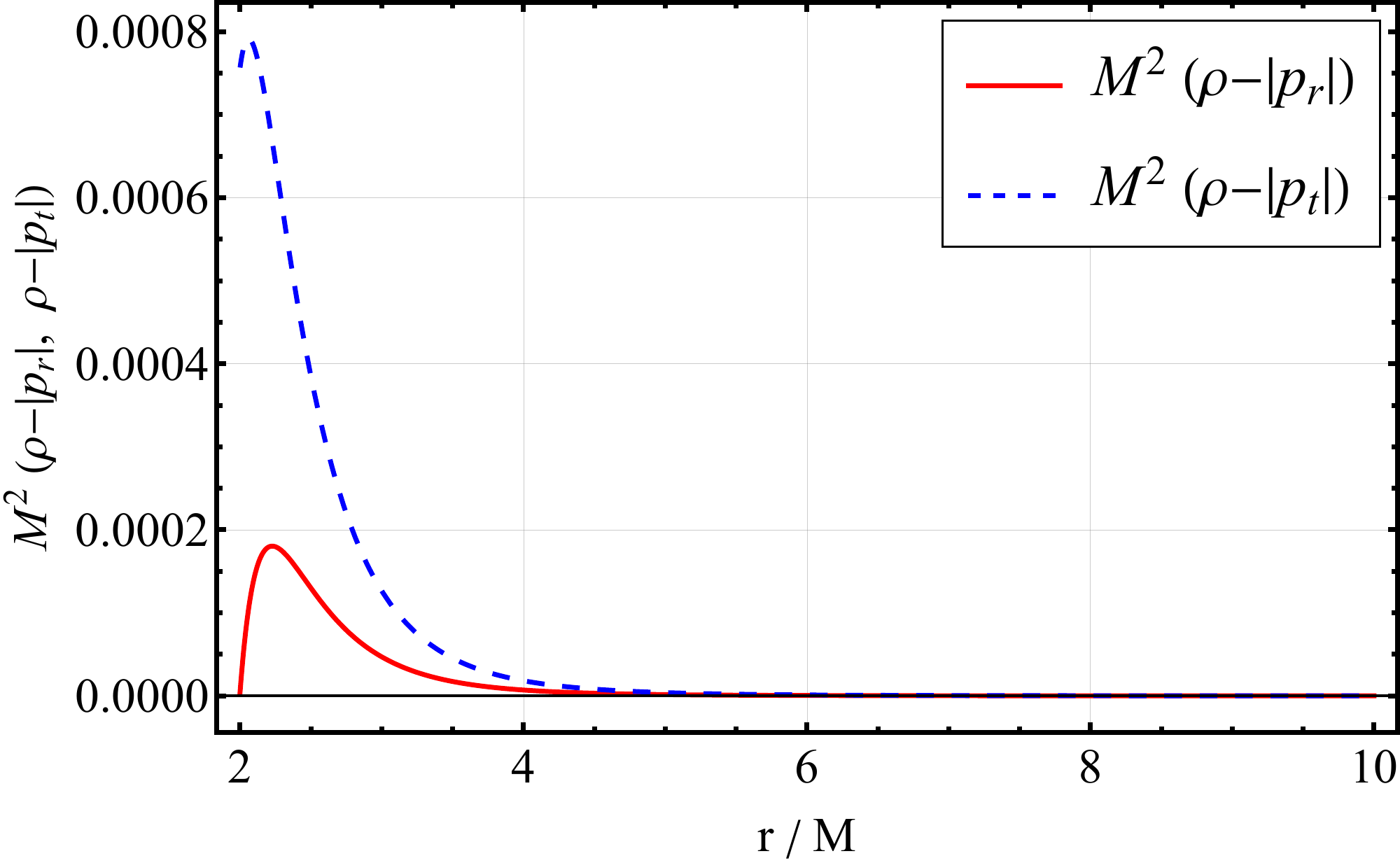}
    \caption{Matter components $\rho$, $p_r$ and $p_t$ (left panel), combinations $\rho+p_r$, $\rho+p_t$, and $\rho+p_r+2p_t$ (right panel) and combinations $\rho-|p_r|$ and $\rho-|p_r|$ (middle panel) as functions of the normalized radial coordinate $r/M$ with $\alpha=5$, $\beta=4$, $\lambda=-100\pi$, $r_0=2M$, and $\zeta_0=-0.1$. Since $\rho+p_r>0$ and $\rho+p_t>0$, one verifies that the NEC is satisfied for the whole spacetime. Furthermore, since $\rho>0$ and $\rho+p_r+2p_t>0$, the WEC and SEC are also satisfied. Finally, since $\rho-|p_r|>0$ and $\rho-|p_r|>0$, the DEC is also satisfied.}
    \label{fig:solution3}
\end{figure*}

\textit{Solution 2: } Let us now consider the restrictions obtained in Eq.\eqref{eq:DECeverywhere}. Under these restrictions, the simplest possible choice is $\beta=4$ and $\alpha=5$. Again, let us also take $r_0=2M$ for some constant with units of mass. From Eq.\eqref{eq:DECboundlambda}, one verifies that for this choice of parameters the constant $\lambda$ must satisfy the restriction $\lambda<-36\pi$. We can thus choose e.g. $\lambda=-100\pi$ to satisfy this constraint. For this combination of parameters, we obtain $\zeta_c\sim -0.240$ and $\zeta_+\sim 0.095$. It is not necessary to verify what constraints arise from $\bar\zeta_\pm$ and $\hat\zeta_\pm$ as we have already proven that these constraints are weaker than the ones arising from $\zeta_\pm$. Finally, one verifies that $\tilde\zeta_c\sim -0.102$ and $\tilde\zeta_c\sim -0.026$, showing that the bounds arising from the DEC are stronger than the ones arising from the NEC, as anticipated. One can thus choose $\zeta_0=-0.1$ to complete the solution. The solutions for the matter variables $\rho$, $p_r$ and $p_t$, as well as the combinations $\rho+p_r$, $\rho+p_t$, $\rho+p_r+2p_t$, $\rho-|p_r|$ and $\rho-|p_t|$, necessary for the NEC, WEC, SEC, and DEC, are plotted in Fig.\ref{fig:solution3}. One can verify that all four energy conditions considered are satisfied for the entire range of the radial coordinate without the necessity of performing a matching with an exterior vacuum spacetime.

\section{Wormhole solutions requiring an exterior matching}\label{sec:matching}

Let us now consider a situation for which a given wormhole solution satisfies the requirements in Eq.\eqref{eq:rest1}, but violates one of the requirements in Eq.\eqref{eq:everywhere}. In this case, the solution
will satisfy the NEC (and possibly also the WEC and SEC) at the throat and up to some critical radius $r_c$, but they will be violated in some subset of the region $r>r_c$. When such a situation arises, a physically relevant wormhole solution can still be constructed via the matching of the interior wormhole solution to an exterior vacuum solution in the region where the energy conditions are still satisfied. For this purpose, one must first determine the junction conditions of the particular case of $f(R,T)$ gravity considered here, which will then be used to perform the matching.

\subsection{Notation and assumptions}

Let us start by introducing the notation to be used in the following sections. Consider a spacetime $\Omega$ which can be decomposed into two distinct regions $\Omega^\pm$ separated by a hypersurface $\Sigma$ of constant radius $r_\Sigma<r_c$, where $r_c$ is the critical radius above which the energy conditions are violated by the interior wormhole solution. Each of the spacetime regions $\Omega^\pm$ is described by a metric $g_{\mu\nu}^\pm$ written in two coordinate systems $x^\mu_\pm$. At the hypersurface $\Sigma$ one defines a set of coordinates $y^a$, where the Latin indices exclude the direction perpendicular to $\Sigma$. The projection vectors from $\Omega^\pm$ to $\Sigma$ are given by $e^\mu_a=\partial x^\mu/\partial y^a$, and we define the normal vector to $\Sigma$ as $n^\mu$, pointing in the direction from $\Omega^-$ to $\Omega^+$. Since the normal vector is a spacelike vector, it satisfies the normalization condition $n_\mu n^\mu=1$ and the orthogonality condition $e^\mu_a n_\mu=0$. One can now define the induced metric $h_{ab}$ and the extrinsic curvature $K_{ab}$ of the hypersurface $\Sigma$ as
\begin{equation}\label{eq:defhab}
h_{ab}=g_{\mu\nu}e^\mu_ae^\nu_b,
\end{equation}
\begin{equation}\label{eq:defKab}
h_{ab}=e^\mu_a e^\nu_b\nabla_\mu n_\nu.
\end{equation}
The affine parameter along the geodesic congruence orthogonal to $\Sigma$ is denoted as $l$ and adequately set to be $l<0$ in the region $\Omega^-$, $l>0$ in the region $\Omega^+$, and $l=0$ at $\Sigma$.

The most appropriate mathematical framework for analysing the junction conditions is the distribution formalism. In this formalism, any regular quantity $X$ can be written in terms of distribution functions as $X=X^+\Theta\left(l\right)+X^-\Theta\left(-l\right)$, where the superscripts $\pm$ indicate the restriction of the quantity $X$ to the regions $\Omega^\pm$ and $\Theta\left(l\right)$ is the Heaviside distribution function, which takes the values $\Theta\left(l<0\right)=0$, $\Theta\left(l>0\right)=1$, and $\Theta\left(0\right)=\frac{1}{2}$. Taking a derivative of such a quantity $X$, one obtains $\partial_\mu X=\partial_\mu X^+\Theta\left(l\right)+\partial_\mu X^-\Theta\left(-l\right)+n_\mu \left[X\right]\delta\left(l\right)$, where $\left[X\right]=X^+|_\Sigma-X^-|_\Sigma$ is the jump of $X$ across $\Sigma$, and $\delta\left(l\right)=\partial_l\Theta\left(l\right)$ is the Dirac delta distribution. Note that, by construction, one has $\left[n^\mu\right]=\left[e_a^\mu\right]=0$.

\subsection{Junction conditions of $f\left(R,T\right)=R+\lambda T$}

Let us now proceed to the derivation of the first junction condition. The junction conditions for a general $f\left(R,T\right)$ gravity were previously obtained in Ref.\cite{rosa3}. However, the junction conditions for the particular case studied in this work, i.e. $f\left(R,T\right)=R+\lambda T$, cannot be obtained simply as a limiting case of the general set, and thus we will briefly derive the adequate set in this section. We start by writing the metric $g_{\mu\nu}$ in the distribution formalism as
\begin{equation}\label{eq:distgab}
g_{\mu\nu}=g_{\mu\nu}^+\Theta(l)+g_{\mu\nu}^-\Theta(-l).
\end{equation}
One must now construct all the necessary geometrical quantities, i.e., the Christoffel symbols, the Riemann tensor, the Ricci tensor, and the Ricci scalar, from the metric above. Taking the derivative of Eq.\eqref{eq:distgab}, one obtains $\partial_\gamma g_{\mu\nu}=\partial_\gamma g^+_{\mu\nu}\Theta(l)+\partial_\gamma g^-_{\mu\nu}\Theta(-l)+[g_{\mu\nu}]n_\gamma\delta(l)$. This result implies that the Christoffel symbols will feature a term proportional to $\delta\left(l\right)$ in the distribution formalism. Although this term is not problematic by itself, the presence of a $\delta\left(l\right)$ term in the Christoffel symbols will give rise to terms proportional to $\delta^2\left(l\right)$ in the Riemann tensor, which are singular in the distribution formalism. To avoid these problematic terms, one must impose a restriction on the continuity of the metric $g_{\mu\nu}$, i.e. $\left[g_{\mu\nu}\right]=0$. This latter condition can be restated into a covariant coordinate-independent form by taking the projection onto $\Sigma$ and using $\left[e^\mu_a\right]=0$, from which one obtains the first junction condition as
\begin{equation}\label{eq:geom_JC1}
    \left[h_{ab}\right]=0.
\end{equation}

Following the junction condition obtained in Eq.\eqref{eq:geom_JC1}, the Riemann tensor and its contractions become regular in the distribution formalism. The Ricci tensor $R_{\mu\nu}$ and the Ricci scalar $R$ can then be written as
\begin{equation}\label{eq:R_tensor_distribution}
\begin{split}
    R_{\mu\nu}=&R_{\mu\nu}^{+} \Theta(l)+R_{\mu\nu}^{-} \Theta(-l)-\\
    &-\left(e^a_\mu e^b_\nu\left[K_{ab}\right]+n_\mu n_\nu [K]\right)\delta(l),
\end{split}
\end{equation}
\begin{equation}\label{eq:R_scalar_distribution}
    R=R^{+} \Theta(l)+R^{-} \Theta(-l)-2[K]\delta(l),
\end{equation}
where $K_{ab}$ is the extrinsic curvature of the hypersurface $\Sigma$ and $K$ is the corresponding trace. The terms proportional to $\delta\left(l\right)$ in Eqs.\eqref{eq:R_tensor_distribution} and \eqref{eq:R_scalar_distribution} can then be associated with the presence of a thin-shell of matter at $\Sigma$. This implies that the stress-energy tensor $T_{\mu\nu}$ must also feature a term proportional to $\delta\left(l\right)$ in the distribution formalism, i.e. we write
\begin{equation}\label{eq:T_tensor_distribution}
T_{\mu\nu}=T_{\mu\nu}^{+} \Theta(l)+T_{\mu\nu}^{-} \Theta(-l)+S_{\mu\nu}\delta(l),
\end{equation}
\begin{equation}\label{eq:T_scalar_distribution}
T=T^{+} \Theta(l)+T^{-} \Theta(-l)+S\delta(l),
\end{equation}
where $S_{\mu\nu}=S_{ab}e^a_\mu e^b_\nu$, $S_{ab}$ is the stress-energy tensor of the thin-shell, and $S=g^{\mu\nu}S_{\mu\nu}=h^{ab}S_{ab}$. For the perfect-fluid case, one has ${S^a}_b=\text{diag}\left(-\sigma,p,p\right)$, where $\sigma$ is the surface energy density and $p$ is the surface pressure of the thin-shell.

Replacing Eqs.\eqref{eq:T_tensor_distribution}, \eqref{eq:R_tensor_distribution}, \eqref{eq:R_scalar_distribution} and \eqref{eq:T_scalar_distribution} into the field equations in Eq. \eqref{eq:fieldeq} and projecting the result onto $\Sigma$ using $e^\mu_a e^\nu_b$, the terms not proportional to $\delta\left(l\right)$ cancel identically and one obtains the second junction condition as
\begin{equation}\label{eq:geom_JC2}
    \left(8\pi+\lambda\right)S_{ab}+\frac{\lambda}{2}S h_{ab}=[K]h_{ab}-\left[K_{ab}\right].
\end{equation}
With this all remaining problematic terms are eliminated and thus Eqs. \eqref{eq:geom_JC2} and \eqref{eq:geom_JC1} constitute the full set of junction conditions of this particular case of $f(R,T)$ gravity.

\subsection{Solutions supported by thin-shells}

We now have all the necessary tools to perform the matching and find wormhole solutions supported by thin-shells that satisfy the NEC for the whole spacetime. The interior region, i.e. $\Omega^-$, is described by the metric given in Eq.\eqref{eq:metric}, whereas for the exterior region, i.e. $\Omega^+$, we consider a Schwarzschild solution described by the line element
\begin{equation}\label{eq:metricext}
    ds^2=-\left(1-\frac{2M}{r}\right)e^{\zeta_e}dt^2+\left(1-\frac{2M}{r}\right)^{-1}dr^2+r^2d\Omega^2,
\end{equation}
where $\zeta_e$ is a constant to be determined in what follows and $M$ represents the mass of the Schwarzschild spacetime. Note that since the Schwarzschild solution is a vacuum solution, one has $T_{\mu\nu}=\Theta_{\mu\nu}=0$ and the field equations in Eq.\eqref{eq:geo_fieldeq} reduce to Einstein's vacuum field equations, which admit the metric in Eq.\eqref{eq:metricext} as a solution. Using Eqs.\eqref{eq:metric} and \eqref{eq:metricext} for $g_{\mu\nu}^-$ and $g_{\mu\nu}^+$ respectively, the first junction condition in Eq.\eqref{eq:geom_JC1} becomes
\begin{equation}\label{eq:JC1_explicit}
    \left(1-\frac{2M}{r_\Sigma}\right)e^{\zeta_e}=e^{\zeta(r_\Sigma)}.
\end{equation}
Once the free parameters of both the interior and the exterior solutions are set, as well as the matching radius $r_\Sigma$, Eq.\eqref{eq:JC1_explicit} can then be used to determine the value of $\zeta_e$.

Let us now turn to the second junction condition in Eq.\eqref{eq:geom_JC2}. To evaluate this condition explicitly, we must determine the extrinsic curvature $K_{ab}$ of the hypersurface $\Sigma$. The extrinsic curvature as seen from the interior ($K_{ab}^-$) and from the exterior ($K_{ab}^+$) spacetimes takes the forms
\begin{equation}\label{eq:extrinsic_wormhole}
    K^-_{ab}=\sqrt{1-\frac{b(r_\Sigma)}{r_\Sigma}}\begin{pmatrix}
    -\frac{\zeta'(r_\Sigma)}{2} & 0 & 0\\
    0 & r_\Sigma & 0\\
    0 & 0 & r_\Sigma\sin^2\theta\end{pmatrix},
\end{equation}
\begin{equation}\label{eq:extrinsic_Schwarzschild}
    K^+_{ab}=\sqrt{1-\frac{2M}{r_\Sigma}}\begin{pmatrix}
    -\frac{M}{{r_\Sigma}^2\left(1-\frac{2M}{r_\Sigma}\right)} & 0 & 0\\
    0 & r_\Sigma & 0\\
    0 & 0 & r_\Sigma\sin^2\theta\end{pmatrix},
\end{equation}
and the corresponding traces are
\begin{equation}\label{eq:extrinsic_trace_wormhole}
    K^-=\frac{4+r_\Sigma\zeta'(r_\Sigma)}{2r_\Sigma}\sqrt{1-\frac{b(r_\Sigma)}{r_\Sigma}},
\end{equation}
\begin{equation}\label{eq:extrinsic_trace_Schwarzschild}
    K^+=\frac{2r_\Sigma-3M}{{r_\Sigma}^2\sqrt{1-\frac{2M}{r_\Sigma}}}.
\end{equation}
Given the spherical symmetry of the system, Eq.\eqref{eq:geom_JC2} will feature only two linearly independent equations, one for the time component and another for the angular components. These equations can be solved for the surface energy density $\sigma$ and surface pressure $p$ of the thin-shell, which take the forms
\begin{equation}\label{eq:shell_energy}
\begin{split}
    \sigma=&-\frac{C}{{r_\Sigma}^2}\left\{\frac{(\lambda+16\pi)r_\Sigma-(\lambda+32\pi)M}{\sqrt{1-\frac{2M}{r_\Sigma}}}-\right.\\
   &-\left.r_\Sigma\left[2\lambda + 32\pi + \lambda r_\Sigma \zeta '(r_\Sigma)\right]\sqrt{1-\frac{b(r_\Sigma)}{r_\Sigma}}\right\},
\end{split}
\end{equation}
\begin{equation}\label{eq:shell_pressure}
\begin{split}
    p=&\frac{C}{{r_\Sigma}^2}\left\{\frac{(3 \lambda + 16 \pi)r_\Sigma-(5 \lambda + 16 \pi)M}{\sqrt{1-\frac{2M}{r_\Sigma}}}-\right.\\
    &-\left.r_\Sigma\left[3\lambda+16\pi-\left(\frac{\lambda}{2} + 8 \pi\right) r_\Sigma \zeta'(r_\Sigma)\right]\sqrt{1-\frac{b(r_\Sigma)}{r_\Sigma}}\right\},
\end{split}
\end{equation}
where we have defined a constant $C\equiv \left[(\lambda-16 \pi) (\lambda+8 \pi)\right]^{-1}$ to simplify the notation.

\begin{figure*}[h]
    \centering
    \includegraphics[scale=0.4]{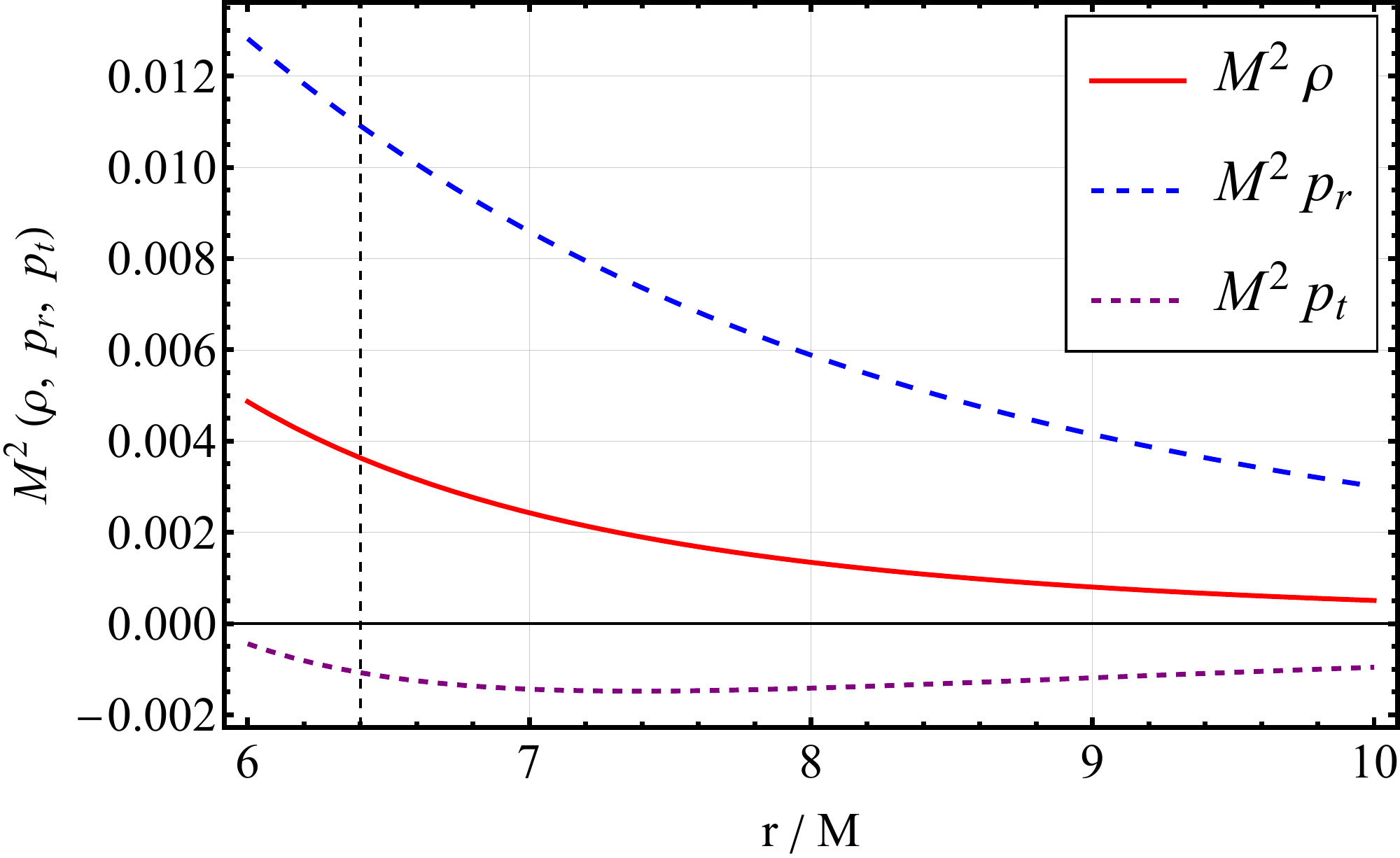}
    \includegraphics[scale=0.4]{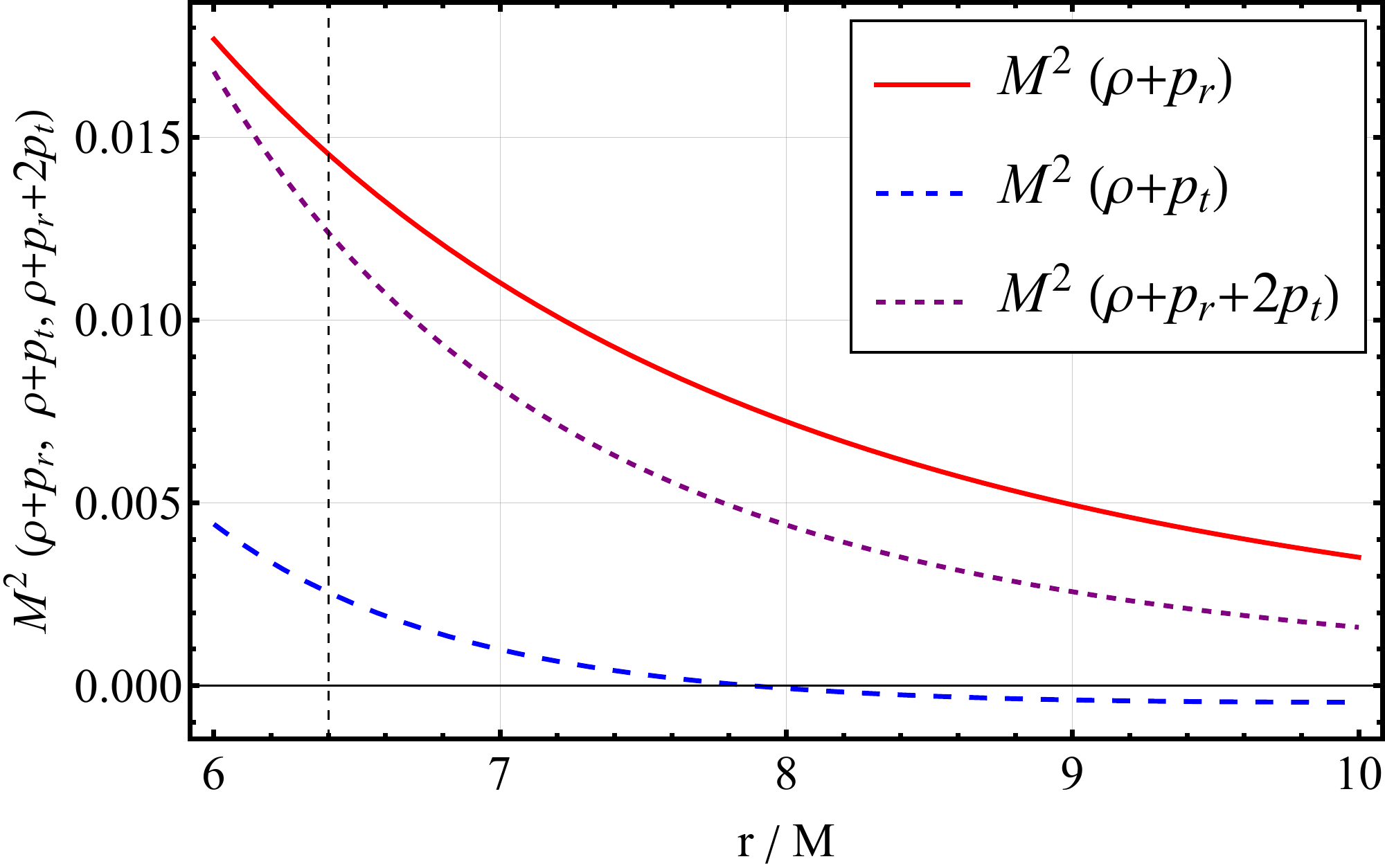}
    \includegraphics[scale=0.4]{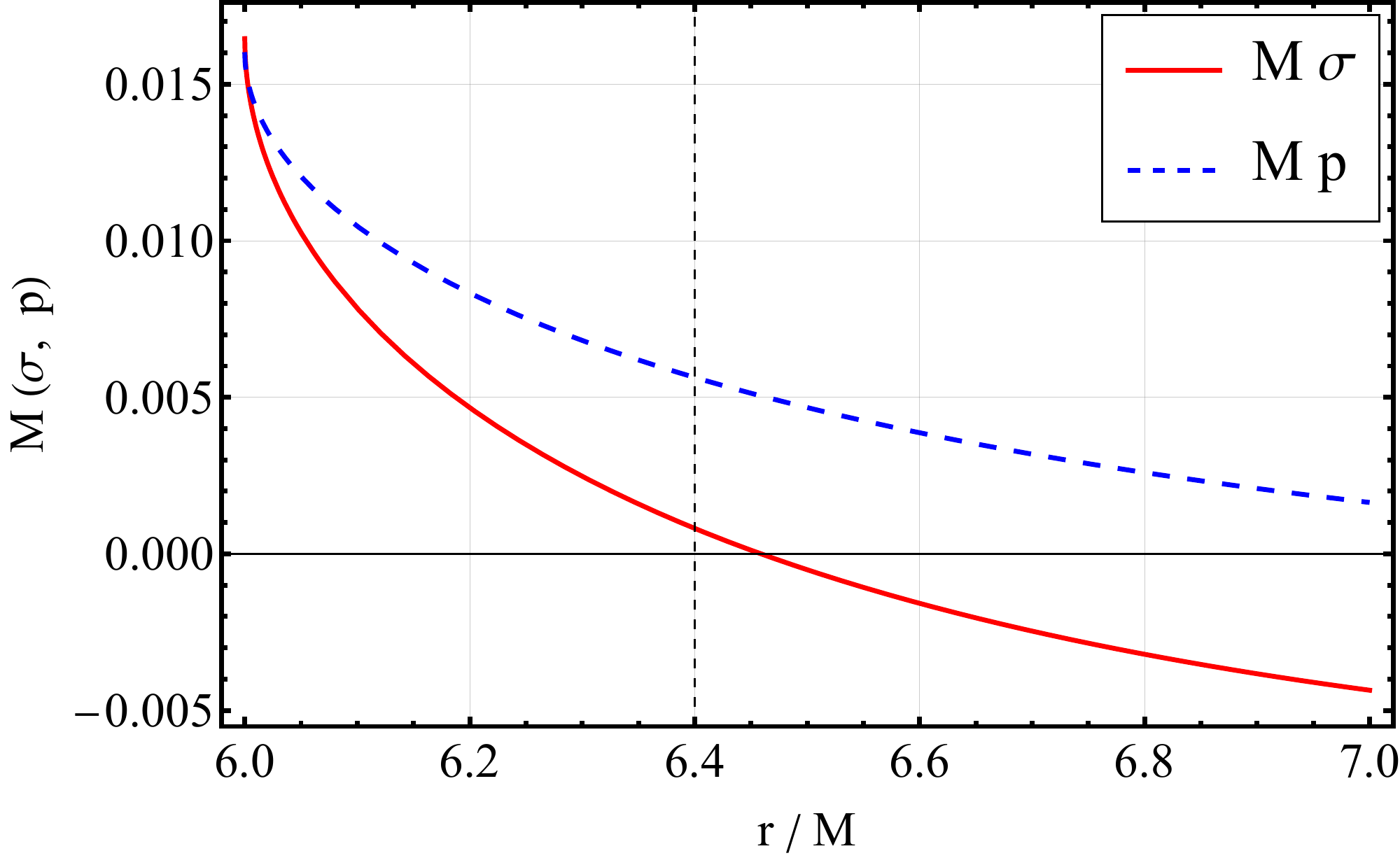}
    \includegraphics[scale=0.4]{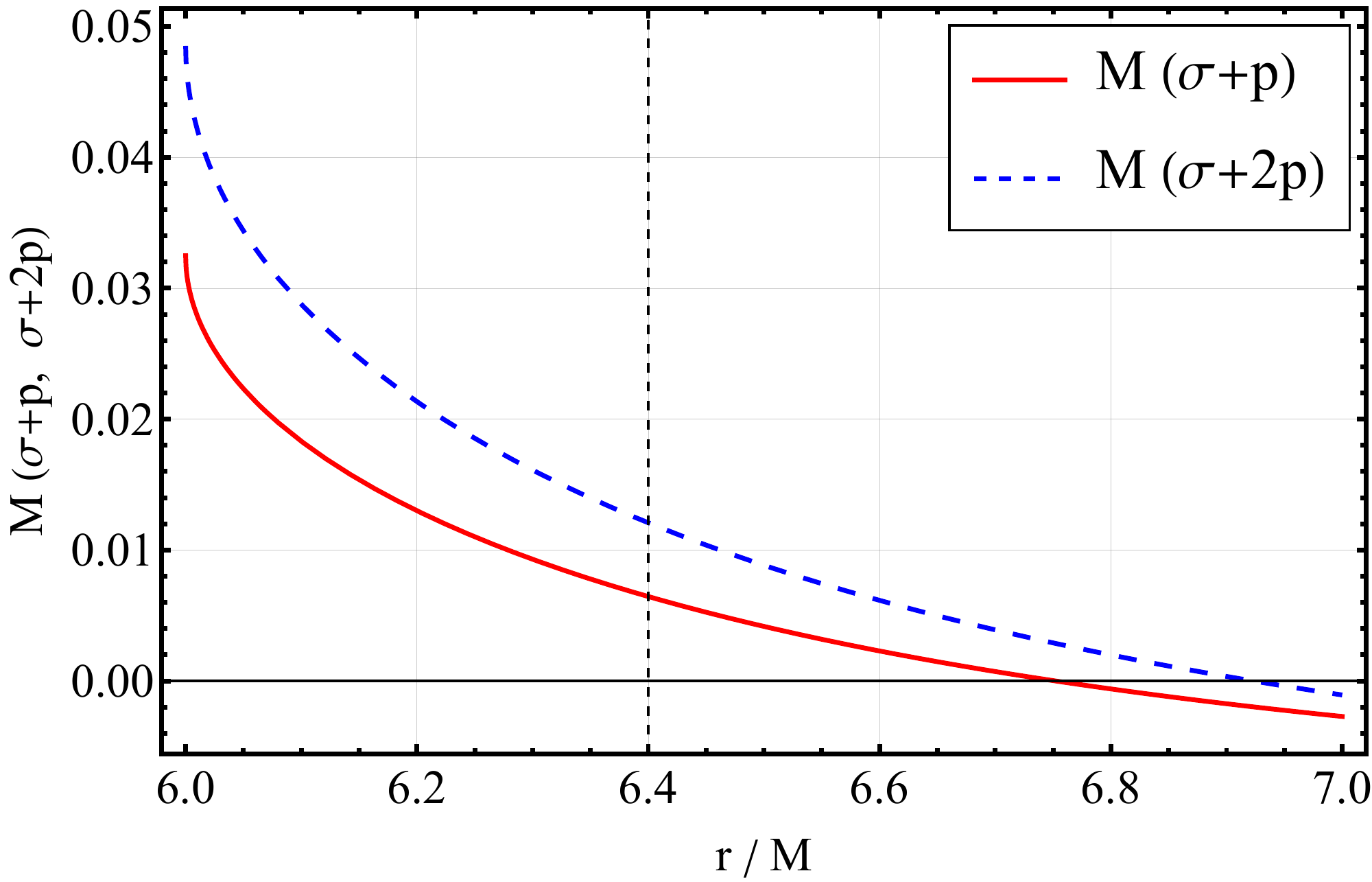}
    \caption{Matter components $\rho$, $p_r$ and $p_t$ (upper left panel), combinations $\rho+p_r$, $\rho+p_t$ and $\rho+p_r+2p_t$ (upper right panel) of the interior (wormhole) spacetime, matter components $\sigma$ and $p$ of the thin-shell at $\Sigma$ (lower left panel) and combinations $\sigma+p$ and $\sigma+2p$ (lower right panel) as functions of the normalized radial coordinate $r/M$ with $\zeta_0=\alpha=\beta=1$, $r_0=6M$ and $\lambda=-9\pi$. Performing the matching at $r_\Sigma=6.4M$, one verifies that the NEC, WEC and SEC are all satisfied for both the interior and the exterior spacetimes, as well as at the separation hypersurface $\Sigma$.}
    \label{fig:solution2}
\end{figure*}

Unlike in the previous section, an analytical study of the energy conditions is impractical due to the complexity of the shell matter quantities given by Eqs.\eqref{eq:shell_energy} and \eqref{eq:shell_pressure}. Instead, and as the purpose of this section is to show that physically relevant solutions can still be constructed when the bounds of the previous section are violated, we will simply provide an explicit example. Consider a combination of parameters with $\zeta_0=\alpha=\beta=1$, $r_0=6M$ and $\lambda=-9\pi$. This combination satisfies the restrictions in Eq.\eqref{eq:rest1}, but violates some of the restrictions in Eq.\eqref{eq:everywhere}, namely $\beta>1$ and $\alpha>\beta+1$, implying that the interior wormhole solution satisfies the NEC only in a finite region of space around the throat $r_0<r<r_c$. One must thus perform a matching with the exterior vacuum spacetime at some $r_\Sigma$ in the same range. In Fig.\ref{fig:solution2} we plot the matter quantities $\rho$, $p_r$, $p_t$, $\sigma$, and $p$, as well as the combinations $\rho+p_r$, $\rho+p_t$, $\rho+p_r+2p_t$ $\sigma+p$ and $\sigma+2p$, for the solution considered. Indeed, one verifies that the NEC is violated in the region $r>r_c\sim 8 M$. Choosing as an example a matching radius of $r_\Sigma=6.4M$, which consequently sets $\zeta_e\sim1.31$ from Eq.\eqref{eq:JC1_explicit}, one effectively removes the region where the NEC is violated from the solution. For this matching radius, one further verifies that the NEC, WEC, and SEC are all satisfied both in the interior spacetime and at the thin-shell, thus providing a strongly physically relevant wormhole solution.

\section{Conclusions}\label{sec:concl}

In this work we have performed a fully analytical parameter space study for a family of wormhole solutions in the linear $f\left(R,T\right)=R+\lambda T$ theory of gravity and obtained the necessary restrictions one must impose on the free parameters of the model in order to guarantee that the wormhole solutions are traversable and non-exotic, i.e., they satisfy all the energy conditions for the entire spacetime. Furthermore, even if some of the parameter bounds are violated and the wormhole becomes exotic outside the throat at some finite radius $r_c$, we have shown that the exoticity can be effectively removed by performing a spacetime matching to an exterior vacuum solution. The wide parameter bounds derived and the flexibility available for a matching with an exterior vacuum indicate that non-exotic wormholes solutions in the $f\left(R,T\right)$ theory are plentiful and that no fine-tuning is required in the search for physically relevant solutions.

For the family of wormholes considered with redshift and shape functions given by the expressions in Eq.\eqref{eq:def_zeta} and \eqref{eq:def_shape}, respectively, and for a linear version of $f\left(R,T\right)=R+\lambda T$, we have shown that forcing the solution to satisfy the NEC for the whole spacetime automatically implies that the solution will also satisfy the WEC and SEC, as the parameter bounds arising from $\rho>0$ and $\rho+p_r+2p_t>0$ are always weaker than the ones arising from $\rho+p_i>0$. Note however that these are one-directional implications, and thus a solution that satisfies $\rho>0$ or $\rho+p_r+2p_t>0$ does not necessarily satisfy the NEC for the whole spacetime. The situation changes for the DEC, where the bounds arising from $\rho>|p_i|$ are effectively stronger than the ones arising from $\rho+p_i>0$. Consequently, a solution satisfying the NEC will not necessarily satisfy the DEC. Nevertheless, we have proven that strong solutions satisfying the four energy conditions, namely the NEC, WEC, SEC and DEC, can still be obtained for a wide variety of parameter combinations.

The linear form $R+\lambda T$ chosen allows one to perform the study of the parameter space analytically and to prove that even the simplest possible extension of GR in the framework of $f\left(R,T\right)$ allows one to successfully and easily solve the problem of exotic matter in wormhole spacetimes. Despite the simplicity of this model, the fact that a plethora of non-exotic solutions were found serves as an indication that in more complicated forms of the theory obtained e.g. via the addition of crossed terms $RT$, even more physically interesting solutions could be lurking. Note that the same is not true for solutions requiring an exterior matching, as the junction conditions of the theory become more restrictive as the complexity of the function $f\left(R,T\right)$ increases. A possible drawback could be the necessity to recur to numerical methods to derive these solutions, but it is undeniable that these solutions exist as these more complicated models still feature the linear model as a limit.

\begin{acknowledgments}
JLR was supported by the European Regional Development Fund and the programme Mobilitas Pluss (MOBJD647).
\end{acknowledgments}



\begin{thebibliography}{99}

\bibitem{morris1}

M. S. Morris, and K. S. Thorne, “Wormholes in spacetime and their use for interstellar travel: A tool for teaching general relativity”, Am. J. Phys. \textbf{56}, 395 (1988).

\bibitem{visser1}

M. Visser, \textit{Lorentzian wormholes: From Einstein to Hawking} (Springer-Verlag, New York, 1996).

\bibitem{visser2}

M. Visser, ”Traversable wormholes: Some simple examples”, Phys. Rev. D \textbf{39}, 3182 (1989).

\bibitem{visser3}

M. Visser, ”Traversable wormholes from surgically modified Schwarzschild spacetimes”, Nucl. Phys. B \textbf{328}, 203 (1989).

\bibitem{lemos1}

J. P. S. Lemos, F. S. N. Lobo, and S. Q. Oliveira, “MorrisThorne wormholes with a cosmological constant”, Phys. Rev. D \textbf{68}, 064004 (2003); arXiv:gr-qc/0302049.

\bibitem{agnese1}

A. G. Agnese and M. La Camera, “Wormholes in the Brans-Dicke theory of gravitation”, Phys. Rev. D \textbf{51}, 2011 (1995).

\bibitem{nandi1}

K. K. Nandi, B. Bhattacharjee, S. M. K. Alam, and J. Evans, “Brans-Dicke wormholes in the Jordan and Einstein frames”, Phys. Rev. D \textbf{57}, 823 (1998); arXiv:0906.0181 [gr-qc].

\bibitem{bronnikov1}

K. A. Bronnikov and S. W. Kim, “Possible wormholes in a brane world”, Phys. Rev. D 67, 064027 (2003); arXiv:gr-qc/0212112.

\bibitem{camera1}

M. La Camera, “Wormhole solutions in the RandallSundrum scenario”, Phys. Lett. B 573, \textbf{27} (2003); arXiv:grqc/0306017.

\bibitem{camera2}

M. La Camera, “Wormhole solutions in the RandallSundrum scenario”, Phys. Lett. B 573, \textbf{27} (2003); arXiv:gr-qc/0306017 [gr-qc].

\bibitem{lobo1}

F. S. N. Lobo, “Exotic solutions in general relativity: Traversable wormholes and ‘warp drive’ spacetimes”, in \textit{Classical and Quantum Gravity Research}, editors M. N. Christiansen and T. K. Rasmussen (Nova Science Publishers, 2008), p. 1; arXiv:0710.4474 [gr-qc].

\bibitem{garattini1}

R. Garattini and F. S. N. Lobo, “Self sustained phantom wormholes in semi-classical gravity”, Classical Quantum Gravity \textbf{24}, 2401 (2007); arXiv:gr-qc/0701020.

\bibitem{lobo2}

F. S. N. Lobo, “General class of wormhole geometries in conformal Weyl gravity”, Classical Quantum Gravity \textbf{25}, 175006 (2008); arXiv:0801.4401 [gr-qc].

\bibitem{garattini2}

R. Garattini and F. S. N. Lobo, “Self-sustained traversable wormholes in noncommutative geometry”, Phys. Lett. B \textbf{671}, 146 (2009); arXiv:0811.0919 [gr-qc].

\bibitem{lobo3}

F. S. N. Lobo and M. A. Oliveira, “General class of vacuum Brans-Dicke wormholes”, Phys. Rev. D \textbf{81}, 067501 (2010); arXiv:1001.0995 [gr-qc]

\bibitem{garattini3}

R. Garattini and F. S. N. Lobo, “Self-sustained wormholes in modified dispersion relations”, Phys. Rev. D \textbf{85}, 024043 (2012); arXiv:1111.5729 [gr-qc].

\bibitem{myrzakulov1}

R. Myrzakulov, L. Sebastiani, S. Vagnozzi, S. Zerbini, ”Static spherically symmetric solutions in mimetic gravity: rotation curves \& wormholes”, Class. Quant. Grav. \textbf{33} (2016) 125005.

\bibitem{lobo4}

F. S. N. Lobo (editor), Wormholes, Warp Drives and Energy Conditions, Fundam. Theor. Phys. \textbf{189}, (Springer International Publishing, 2017).

\bibitem{lobo5}

F. S. N. Lobo and M. A. Oliveira, “Wormhole geometries in f(R) modified theories of gravity”, Phys. Rev. D \textbf{80}, 104012 (2009); arXiv:0909.5539 [gr-qc].

\bibitem{garcia1}

N. M. Garcia and F. S. N. Lobo, “Wormhole geometries supported by a nonminimal curvature-matter coupling”, Phys. Rev. D \textbf{82}, 104018 (2010); arXiv:1007.3040 [gr-qc].

\bibitem{garcia2}

N. Montelongo Garcia and F. S. N. Lobo, “Nonminimal curvature-matter coupled wormholes with matter satisfying the null energy condition”, Class. Quant. Grav. \textbf{28}, 085018 (2011); arXiv:1012.2443 [gr-qc].

\bibitem{harko1}

T. Harko, F. S. N. Lobo, M. K. Mak, and S. V. Sushkov, “Modified-gravity wormholes without exotic matter”, Phys. Rev. D \textbf{87}, 067504 (2013); arXiv:1301.6878 [gr-qc].

\bibitem{bhawal1}

B. Bhawal and S. Kar, “Lorentzian wormholes in Einstein–Gauss-Bonnet theory”, Phys. Rev. D \textbf{46}, 2464 (1992).

\bibitem{dotti1}

G. Dotti, J. Oliva, and R. Troncoso, “Exact solutions for the Einstein-Gauss-Bonnet theory in five dimensions: Black holes, wormholes and spacetime horns”, Phys. Rev. D \textbf{75}, 024002 (2007); arXiv:0706.1830 [hep-th].

\bibitem{mehdizadeh1}

M. R. Mehdizadeh, M. Kord Zangeneh, and F. S. N. Lobo, “Einstein-Gauss-Bonnet traversable wormholes satisfying the weak energy condition”, Phys. Rev. D \textbf{91},
no. 8, 084004 (2015); arXiv:1501.04773 [gr-qc].

\bibitem{anchordoqui1}

L. A. Anchordoqui, S. E. Perez Bergliaffa, and D. F. Torres, “Brans-Dicke wormholes in nonvacuum spacetime”, Phys. Rev. D \textbf{55}, 5226 (1997); arXiv:gr-qc/9610070.

\bibitem{lobo6}

F. S. N. Lobo, “A general class of braneworld wormholes”, Phys. Rev. D \textbf{75}, 064027 (2007); arXiv:grqc/0701133.

\bibitem{capozziello1}

S. Capozziello, T. Harko, T. S. Koivisto, F. S. N. Lobo, and G. J. Olmo, “Wormholes supported by hybrid metric-Palatini gravity”, Phys. Rev. D \textbf{86}, 127504 (2012) [arXiv:1209.5862 [gr-qc]].

\bibitem{rosa1}

J.~L.~Rosa, J.~P.~S.~Lemos and F.~S.~N.~Lobo, ``Wormholes in generalized hybrid metric-Palatini gravity obeying the matter null energy condition everywhere,'' Phys. Rev. D \textbf{98} (2018) no.6, 064054 [arXiv:1808.08975 [gr-qc]].

\bibitem{rosa2}

J.~L.~Rosa, ``Double gravitational layer traversable wormholes in hybrid metric-Palatini gravity,'' Phys. Rev. D \textbf{104} (2021) no.6, 064002 [arXiv:2107.14225 [gr-qc]].

\bibitem{rosalol}

J.~L.~Rosa and J.~P.~S.~Lemos, ``Junction conditions for generalized hybrid metric-Palatini gravity with applications,'' Phys. Rev. D \textbf{104} (2021) no.12, 124076 [arXiv:2111.12109 [gr-qc]].

\bibitem{harko2}

T.~Harko, F.~S.~N.~Lobo, S.~Nojiri and S.~D.~Odintsov, "$f(R,T)$ gravity," Phys. Rev. D \textbf{84} (2011), 024020 [arXiv:1104.2669 [gr-qc]].

\bibitem{zaregonbadi1}

R. Zaregonbadi, M. Farhoudi and N. Riazi. “Dark matter from $f(R,T)$ gravity”, Phys. Rev. D \textbf{94} (2016) 084052.

\bibitem{dey1}

S. Dey, A. Chanda and B. C. Paul. “Compact objects in $f(R,T)$ gravity with Finch–Skea geometry”, EPJPlus \textbf{136} (2021) 2 228.

\bibitem{carvalho1}

G. A. Carvalho, R. V. Lobato, P. H. R. S. Moraes, J. D. V. Arba˜nil, E. Otoniel, R. M. Marinho Jr, M. Malheiro, ”Stellar equilibrium configurations of white dwarfs in the $f (R, T)$ gravity”, The European Physical Journal C volume \textbf{77}, 871 (2017)

\bibitem{deb1}

D. Deb, F. Rahaman, S. Ray, B.K. Guha, ”Strange stars in $f (R, T)$ gravity”, JCAP\textbf{03} 044 (2018).

\bibitem{maurya1}

S.K. Maurya, A. Errehymy, D. Deb, F. Tello-Ortiz, M. Daoud, ”Study of anisotropic strange stars in f (R, T) gravity: An embedding approach under the simplest linear functional of the matter-geometry coupling”, Phys. Rev. D \textbf{100}, 044014 (2019).

\bibitem{bhatti1}

M. Z. Bhatti, Z. Yousaf, M. Yousaf, ”Stability of self-gravitating anisotropic fluids in $f (R, T)$ gravity”, Physics of the Dark Universe \textbf{28} 100501 (2020).

\bibitem{velten1}

H. Velten and T. R. P. Caramˆes. “Cosmological inviability of $f(R,T)$ gravity”, Phys. Rev. D \textbf{95} (2017) 123536.

\bibitem{mirza1}

B.~Mirza and F.~Oboudiat, ``A Dynamical System Analysis of $f(R,T)$ Gravity,'' Int. J. Geom. Meth. Mod. Phys. \textbf{13} (2016) no.9, 1650108 [arXiv:1412.6640 [gr-qc]].

\bibitem{houndjo1}

M. J. S. Houndjo, “ Reconstruction of $f(R, T)$ gravity describing matter dominated and accelerated phases”, Int. J. Mod. Phys. D. \textbf{21}, 1250003 (2012).

\bibitem{houndjo2}

M. J. S. Houndjo and O. F. Piattella, “Reconstructing $f(R, T)$ gravity from holographic dark energy”, Int. J. Mod. Phys. D. \textbf{21}, 1250024 (2012).

\bibitem{jamil1}

M. Jamil, D. Momeni, M. Reza and R. Myrzakulov, ”Reconstruction of some cosmological models in $f (R, T)$ cosmology”, Euro. Phys. J. C 72, 1999 (2012)

\bibitem{alvarenga1}

F. G. Alvarenga, M. J. S. Houndjo, A. V. Monwanou, J. B. C. Orou, ”Testing some $f(R,T)$ gravity models from energy conditions”, Journal of Modern Physics \textbf{4}, 130-139 (2013).

\bibitem{wu1}

J. Wu, G. Li, T. Harko, S. D. Liang, ”Palatini formulation of $f (R, T)$ gravity theory, and its cosmological implications”, Euro. Phys. J. C 78, 430 (2018).

\bibitem{rosa3}

J.~L.~Rosa, ``Junction conditions and thin shells in perfect-fluid $f(R,T)$ gravity,'' Phys. Rev. D \textbf{103} (2021) no.10, 104069 [arXiv:2103.11698 [gr-qc]].

\bibitem{rosa4}

J.~L.~Rosa and D.~Rubiera-Garcia,

\bibitem{dixit1}

A. Dixit, C. Chawla and A. Pradhan, "Traversable wormholes with logarithmic shape function in $f(R,T)$ gravity", International Journal of Geometric Methods in Modern Physics \textbf{18} (2021) no. 04, 2150064.

\bibitem{banerjee1}

A. Banerjee, M.K. Jasim and S. G. Ghosh. “Wormholes in $f(R,T)$ gravity satisfying the null energy condition with isotropic pressure”, Annals of Physics \textbf{433} (2021) 168575. 

\bibitem{mishra1}

A. K. Mishra, U. K. Sharma, V. C. Dubey, and A. Pradhan, "Traversable wormholes in $f(R,T)$ gravity", Astrophys. Space Sci. \textbf{365} (2020), 34.

\bibitem{sahoo1}

P. Sahoo, P.H.R.S. Moraes, M. M. Lapola, P.K. Sahoo, ”Traversable wormholes in the traceless $f (R, T)$ gravity”, Int. Journ. Mod. Phys. D \textbf{30} (2021) no.13, 2150100, arXiv:2012.00258 [gr-qc].

\bibitem{moraes1}

P. H. R. S. Moraes, P. K. Sahoo, ”Modeling wormholes in $f (R, T)$ gravity”, Phys. Rev. D \textbf{96}, 044038 (2017).

\bibitem{goncalves1}

T.~B.~Gon\c{c}alves, J.~L.~Rosa and F.~S.~N.~Lobo, ``Cosmology in scalar-tensor $f(R,T)$ gravity,'' Phys. Rev. D \textbf{105} (2022) no.6, 064019 [arXiv:2112.02541 [gr-qc]].

\bibitem{goncalves2}

T.~B.~Gon\c{c}alves, J.~L.~Rosa and F.~S.~N.~Lobo, ``Cosmological sudden singularities in $f(R,T)$ gravity,'' Eur. Phys. J. C \textbf{82} (2022) no.5, 418
[arXiv:2203.11124 [gr-qc]].

\bibitem{pinto1}

M. A. S. Pinto, T. Harko,  F. S. N. Lobo, "Gravitationally induced particle production in scalar-tensor $f(R,T)$ gravity", Phys. Rev. D 106, 044043 (2022)

\bibitem{bazeia1}

D.~Bazeia, A.~S.~Lob\~ao and J.~L.~Rosa, ``Multi-kink braneworld configurations in the scalar-tensor representation of f(R,~T) gravity,'' Eur. Phys. J. Plus \textbf{137} (2022) no.9, 999 [arXiv:2209.01928 [gr-qc]].

\bibitem{rosa5}

J.~L.~Rosa, A.~S.~Lob\~ao and D.~Bazeia, ``Impact of compactlike and asymmetric configurations of thick branes on the scalar\textendash{}tensor representation of $f\left( R,T\right) $ gravity,'' Eur. Phys. J. C \textbf{82} (2022) no.3, 191 [arXiv:2202.10713 [gr-qc]].

\bibitem{rosa6}

J.~L.~Rosa, D.~Bazeia and A.~S.~Lob\~ao, ``Effects of Cuscuton dynamics on braneworld configurations in the scalar\textendash{}tensor representation of $f\left( R,T\right) $ gravity,'' Eur. Phys. J. C \textbf{82} (2022) no.3, [arXiv:2111.08089 [gr-qc]].

\bibitem{rosa7}

J.~L.~Rosa, M.~A.~Marques, D.~Bazeia and F.~S.~N.~Lobo, ``Thick branes in the scalar\textendash{}tensor representation of f(R,~T) gravity,'' Eur. Phys. J. C \textbf{81} (2021) no.11, 981 [arXiv:2105.06101 [gr-qc]].

\bibitem{kullthesis}

P. M. Kull, "Radially anisotropic wormholes in $f(R,T)$ gravity", (2022) BSc thesis, University of Tartu. http://hdl.handle.net/10062/83883.





\end{thebibliography}
\end{document}